\documentclass[manuscript]{aastex}
\usepackage{color}
\usepackage{ulem}

\newcommand{\myemail}{akihiro.doi@vsop.isas.jaxa.jp}
\slugcomment{Published in ApJ, 807, 15}

\shorttitle{Core Shift in NGC~4261}
\shortauthors{Haga et al.}

\begin{document}


\title{Determination of Central Engine Position and Accretion Disk Structure in~\objectname{NGC~4261} by Core Shift Measurements}
% \objectname{NGC~4261} 
% core shift measurement 
% the position of the black hole


\author{Takafumi Haga\altaffilmark{1}}
\affil{Department of Space and Astronautical Science, The Graduate University for Advanced Studies~(SOKENDAI), 3-1-1 Yoshinodai, Chuou-ku, Sagamihara, Kanagawa, 252-5210, Japan}

\author{Akihiro  Doi\altaffilmark{2} and Yasuhiro Murata\altaffilmark{2}} 
\affil{The Institute of Space and Astronautical Science, Japan Aerospace Exploration Agency, 3-1-1 Yoshinodai, Chuou-ku, Sagamihara, Kanagawa, 252-5210, Japan; \myemail}

\author{Hiroshi  Sudou\altaffilmark{3}} 
\affil{Department of Mathematical and Design Engineering, Faculty of Engineering, Gifu University, 1-1 Yanagido, Gifu City 5011-1193, Japan}

\author{Seiji Kameno\altaffilmark{4}} 
\affil{Joint ALMA Observatory, Alonso de C$\acute{o}$rdova  3107 Vitacura, Santiago, Chile}

\and

\author{Kazuhiro Hada\altaffilmark{5,6}}
\affil{Mizusawa VLBI Observatory, National Astronomical Observatory of Japan, Osawa, Mitaka, Tokyo 181-8588, Japan}
\affil{INAF Istituto di Radioastronomia, via Gobetti 101, I-40129 Bologna, Italy}


%\altaffiltext{1}{Visiting Astronomer, Cerro Tololo Inter-American Observatory.
%CTIO is operated by AURA, Inc.\ under contract to the National Science
%\altaffiltext{2}{Society of Fellows, Harvard University.}
%\altaffiltext{3}{present address: Center for Astrophysics,
 %   60 Garden Street, Cambridge, MA 02138}


\begin{abstract}
We report multi-frequency phase-referenced observations of the nearby radio galaxy \objectname{NGC~4261}, which has prominent two-sided jets, using the Very Long Baseline Array~(VLBA) at 1.4--43~GHz.  
We measured radio core positions showing observing frequency dependences (known as ``core shift'') in both approaching and counter jets.
The limit of the core position as the frequency approaches infinity, which suggests a jet base, is separated by ($82\pm16 $)~micro-arcsecond upstream in projection, corresponding to ($310\pm60$)~$R_\mathrm{s}$ ($R_\mathrm{s}$; Schwarzschild radius) as a deprojected distance, from the 43~GHz core in the approaching jet.  
In addition, the innermost component at the counter jet side appeared to approach the same position at infinity of the frequency, indicating that cores on both sides approach the same position, suggesting a spatial coincidence with the central engine.  
Applying a phase-referencing technique, we also obtained spectral index maps, 
which indicate that emission from the counter jet is affected by free--free absorption (FFA).
The result of the core shift profile on the counter jet also requires FFA because the core positions at 5--15~GHz cannot be explained by a simple core shift model based on synchrotron self-absorption (SSA).  
Our result is apparently consistent with the SSA core shift with an additional disk-like absorber over the counter jet side.
Core shift and opacity profiles at the counter jet side suggests a two-component accretion: a radiatively inefficient accretion flow (RIAF) at the inner region and a truncated thin disk in the outer region.   
We proposed a possible solution about density and temperature profiles in the outer disk on the basis of the radio observation.  
\end{abstract}

\keywords{galaxies: active --- accretion, accretion disks --- galaxies: jets --- galaxies: supermassive black holes --- galaxies: individual (NGC~4261)}


\section{Introduction}
%%%%%%%%%%%%%%%%%%%%%%%%%%%%
It is widely believed that the release of gravitational energy from an disk of accreting matter onto a super massive black hole (SMBH) generates a huge amount of energy in active galactic nuclei (AGNs).  
Energy and matter can be ejected in the form of jets, which have relativistic velocities and emit non-thermal radiation. 
The jets of radio galaxies, develop from the central engine beyond their host galaxy and are characteristically narrow.
Very long baseline interferometry (VLBI) images of AGNs often reveal a "radio core," a region of peak intensity generated by a compact component at the apparently upstream end of the jets.  

The radio cores of blazars or BL Lacertae objects manifest as standing shock points up to $10^4$--$10^6$~$R_\mathrm{s}$ (Schwarzschild radius) distant from the black hole \citep{Jorstad:2001, Jorstad:2005, Marscher:2008}. 
In other objects, such as the \objectname{M87} galaxy, the central engine is located within a few tens of $R_\mathrm{s}$ from the 43~GHz radio core \citep{Hada:2011}.

The \objectname{M87} result was obtained by measuring core shifts, defined as the positional change of the core at different frequencies caused by absorption of the base of jet. 
The infinite-frequency extrapolation of the core position can be interpreted as the origin of the jet.   
\citet{Hada:2011} estimated the position of the central engine of \objectname{M87} by measuring the core shifts at varying frequency and observing an exterior source as a phase reference.  

The absorption processes that cause core shifts are synchrotron self-absorption (SSA) and free--free absorption (FFA). 
In SSA, the core is located at the position of unity opacity. 
SSA opacity depends on physical conditions of jet, such as the luminosity, structure, velocity, and inclination, etc.~of the jet, as well as the observed emission frequencies.  
Besides \objectname{M87}, this type of core shift has been measured in \objectname{3C~395} \citep{Lara:1994}, \objectname{3C~345} \citep{Lobanov:1998}, \objectname{M81} \citep{Bietenholz:2004}, among others \citep{Sokolovsky:2011,Kovalev:2008}. 

Core shifts arising from FFA potentially provide information on the size, density, and temperature of the absorbing material. 
FFA absorbers existing within sub-pc distances have been suggested in some objects such as \objectname{3C~84} \citep{Vermeulen:1994, Walker:1994},  \objectname{Cen~A} \citep{Tingay:2001},  \objectname{Cyg~A} \citep{Krichbaum:1998} and \objectname{NGC~1052} \citep{Kameno:2001}. Core shifts have also been reported in \objectname{Cyg~A} and \objectname{NGC~1052} \citep{Kadler:2004}.

Measurements of core shifts and studies of the distributions of a spectral index require an unambiguous reference position when comparing core positions at different frequencies and overlaying images collected at different frequencies.  
Previous studies performed properly using optically-thin jet components as reference positions \citep[e.g.,][]{Kovalev:2008,Kameno:2001,Sokolovsky:2011} or an exterior source by phase-referencing technique \citep[e.g.,][]{Hada:2011,Sudou:2003}.  
However, some aspects of core shifts remain elusive. 
% Among these is fuzziness of the connection between the limiting core position and central engine location. 
Although the extrapolation of core positions to infinite frequency is supposed to locate a central engine, it is a working hypothesis.  Only we can claim rigidly is that a central engine and a black hole exist somewhere upstream of observed cores, as far as measurements perform on single-sided jet sources.  
%Whether this fuzziness really exists is not easily determined, because most core shift measurements have been performed on single-sided jet sources. 
Previously, core shift measurements both in approaching and counter jets have been attempted for NGC~1052 under the lack of a common reference position over observing frequencies; however, multiple candidates for the location of central engine are discussed \citep{Kadler:2004}.   
%these are rendered inaccurate because of  appropriately.
%
A core position was frequently assumed as the reference position in previous studies, for example, when spectral index maps were made.  However, as discussed above, this may alter with frequency because of absorption.   
To resolve these problems, we supplement the analysis of a two-sided jet source with phase-referencing observations. \objectname{NGC~4261} is a suitable AGN candidate on which this technique is applied.

%1.2
\objectname{NGC~4261} (3C~270) is a nearby Fanaroff--Riley Class I (FR-I) radio galaxy. 
It is characterized by nearly symmetric kiloparsec-scale two-sided jets \citep{Birkinshaw:1985}. 
The galaxy is 31.6~Mpc distant from the Earth \citep{Tonry:2000}, corresponding to 0.15~pc/mas. 
This proximity presents an advantage for studying regions within 1~pc of the central engine. %because --.
The \objectname{NGC~4261} nucleus is known to contain a central black hole of mass $(4.9 \pm 1.0)\times 10^{8}~M_{\odot}$ \citep{Ferrarese:1996}. 
Hubble Space Telescope (HST) observations have also revealed a disk of gas and dust, having a diameter of ~300 pc, in the \objectname{NGC~4261} nucleus \citep{Jaffe:1993}.

Another feature of \objectname{NGC~4261}, revealed by VLBI studies, is its prominent parsec-scale two-sided jet along large-scale jets, running parallel to the rotation axis of the dust disk \citep{Jones:2000}. 
The West- and East-side jets are approaching and receding to observers, respectively; the latter is known as a counter jet. 
The brightness of both jets is slightly affected by relativistic beaming. 
\citet{Piner:2001} estimated the speed of the jets by very long baseline array (VLBA) monitor observations at 8~GHz. 
From the jet-to-counter jet brightness ratio and apparent speed, they obtained a gentle intrinsic jet velocity of about $(0.46\pm0.02)c$, inclined at $63\pm 3\degr$ from the line of sight.

\objectname{NGC~4261} is further characterized by an apparent discontinuity in the structure of the counter jet. 
This region, called the ``gap'', has been interpreted as obscuration by an edge-on, geometrically thin, cold parsec-scale disk with a temperature of $\sim 10^4$~K \citep{Jones:2001}. 
As an alternative to the FFA disk model, the gap may arise from resistive radiatively inefficient accretion flow \citep{Kaburaki:2010}. 
According to \citet{Kaburaki:2010}, the low temperature of the standard disk model \citep{Shakura:1973} precludes the existence of FFA plasma gas in a gap region so distant from the central black hole.

The present study aims to determine the position of the central engine in \objectname{NGC~4261} and investigate the structure of the absorber. 
In section 2, we explain the phase-referencing method from multi-frequency observations and core shift measurements. 
Section 3 is devoted to core shift results and spectral index maps. 
The position of the central engine, the interpretation of the core shift, the absorption process and the disk model of \objectname{NGC~4261} are discussed in section 4. 
Section 5 concludes the paper.

\section{Observations and data reduction}
\subsection{Observations}
Multi-frequency observations of \objectname{NGC~4261} were conducted using the VLBA \citep{Napier:1994} at four frequency bands (1.4, 2.3, 5.0, and 8.4~GHz) on July 3, 2003, and at three frequency bands (15, 22, and 43~GHz) on June 26, 2003. 
The two sets of observations were separated by only 7~days. 
Each observation period was approximately 12~hours. 
At 1.4~GHz, dual-circular polarizations at two intermediate frequencies (IF) were received, covering a total bandwidth of 16~MHz. 
At both 2.3 and 8.4~GHz, two IFs covered a total bandwidth of 16~MHz in right-hand circular polarization. 
Other frequency bands collected four IFs, covering a total bandwidth of 32~MHz in left-hand circular polarization.  Digital sampling of 2-bit was applied at all observational frequency bands.
Correlations were processed using the VLBA correlator in Socorro, New Mexico.

The radio core positions of \objectname{NGC~4261} were measured by the nodding-style phase-referencing technique, relative to a nearby reference source, J1222+0413, separated by less than $2^\circ$ from \objectname{NGC~4261}; ��-T-C1-T-C1-T-��, where T and C1 represent the target and primary calibrator, respectively. 
Switching between the target and calibrator was repeated after scanning for 20 s. 
Observations were conducted for approximately 20 minutes at each frequency in turn. 
Under this measurement condition, the {\it uv}-coverage among frequencies was almost identical, thereby minimizing the residual systematic errors in astrometry. 
The total on-source time on the target was approximately one hour at each frequency. 

\subsection{Calibration and Imaging}
The antenna parallactic angles and Earth orientation parameters were corrected.  Furthermore, ionospheric dispersive delays were corrected using the ionospheric model from GPS data provided by the Jet Propulsion Laboratory \citep{Mannucci:1999}.  
Instrumental delays/phases and bandpass were calibrated using \objectname{3C~273} data. 
Assuming a point source model, we performed fringe-fitting on J1222+0413 as a phase calibrator C1. 
Subsequently, we calibrated the visibility amplitude and phase using a source structure model available in the Difmap software \citep{Shepherd:1994} procedures. 
This model is based on iterative deconvolution and self-calibration algorithms. 
Applying these solutions to the visibilities of \objectname{NGC~4261}, we obtained phase-referenced images of \objectname{NGC~4261}.  On the other hand, we also made self-calibrated images for NGC~4261 using the Difmap (Figure~\ref{fig:all_cln}).  

\subsection{Core position measurement}\label{section:corepositionmeasurement} 
\begin{figure}
\epsscale{0.75}
\plotone{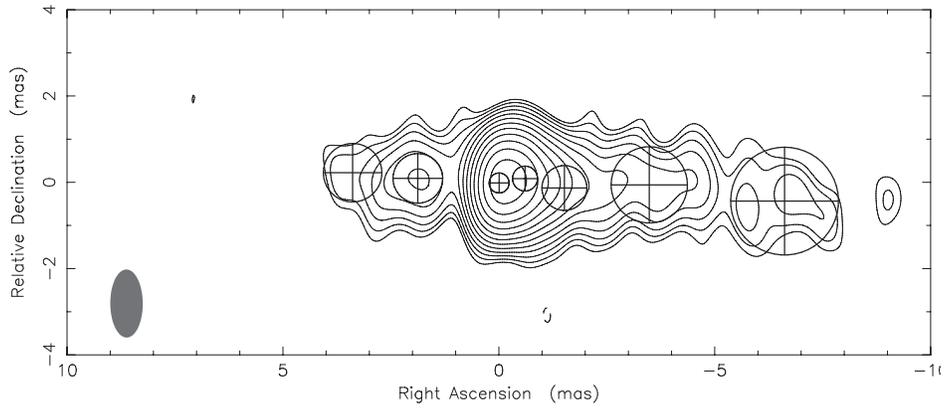}
\caption{Contour Image of NGC~4261 at 8~GHz and overlaid fitted components as an example for circular-Gaussian model fitting using {\tt modelfit} in Difmap.
}
\label{fig:8G}
\end{figure}

The source position in the phase-referenced images was measured using {\tt JMFIT} in AIPS \citep{Greisen:2003}, adopting a single-ellipse Gaussian profile model. 
However, we consider that, because the core and jets were blended, the core position potentially drifted toward the jet side from its true position. 
The jet contribution was evaluated by comparing results by two methods of measuring the core position in final~(self-calibrated) images of NGC~4261 and J1222+0413; one-component {\tt JMFIT} and fitting of visibility data to multi circular Gaussian components using {\tt modelfit} in Difmap (Figure~\ref{fig:8G}). 
% The methods were iteratively performed on self-calibrated data.   
The final core position was defined as the positions between the core in the phase-referenced image of NGC~4261 with the correction of jet contribution on its final image and the core of J1222+0413 in the final images with the correction of jet contribution.  

Following the method of \citet{Hada:2011}, we estimated the error in our astrometric measurements. 
The error incorporates ionospheric residuals, tropospheric residuals, antenna positional error, Earth orientation error, a priori source coordinate error, core identification error, and thermal noise. 
The error contributions are listed in Table~\ref{tab:eb}.
Based on GPS data, we assume a total electron content of $3 \times 10^{17} \rm{m}^{-2}$, which is reducible to 1/4 equivalent following ionospheric residual correction by {\tt TECOR} in AIPS \citep{Mannucci:1999}. 
The zenith excess path error in tropospheric residuals is assumed to be 2~cm \citep{Beasley:1995}. 
We adopted proper geometric errors such as antenna location error (3~mm), Earth orientation parameter error (VLBA memo~69), and a priori source coordinate error that is determined from source declination \citep{Pradel:2006}. 
The core identification error was estimated from the difference in core positions yielded by the above-described two fitting methods. 
The thermal noise error was defined by the beam size over the dynamic range of phase-referenced maps.

\section{Results}
\subsection{Multi-frequency CLEAN images}
\begin{figure}
\epsscale{0.6}
\plotone{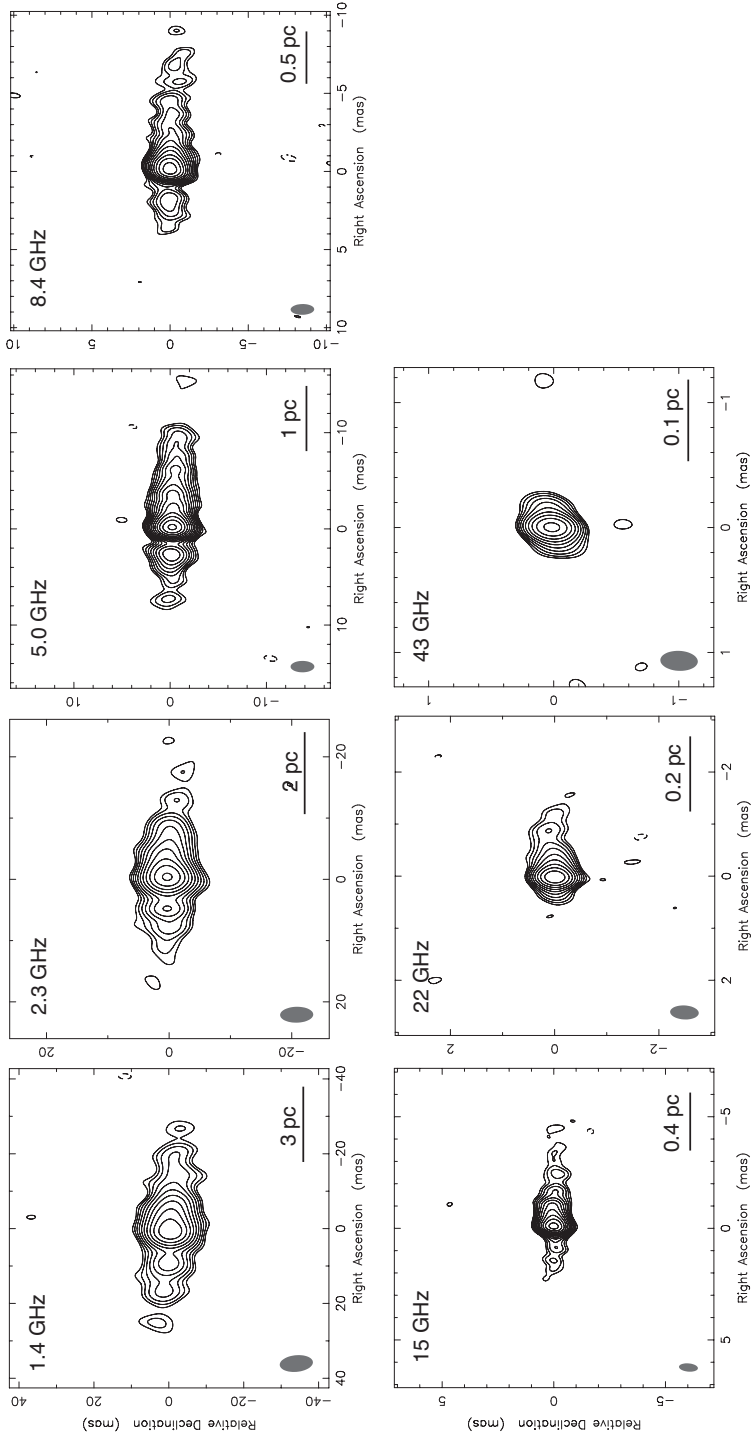}
\caption{Self-calibrated images of NGC~4261 at each frequency from VLBA observations. 
The gray ellipses show the synthesized beam size (FWHM). 
The RMS noise in the images at each frequency is provided in Table~\ref{tab:clean}. 
Contours start at the $\pm 3 \sigma$ level, increasing by factors of $\sqrt 2$.
}
\label{fig:all_cln}
\end{figure}

\begin{figure}
\epsscale{.4}
\plotone{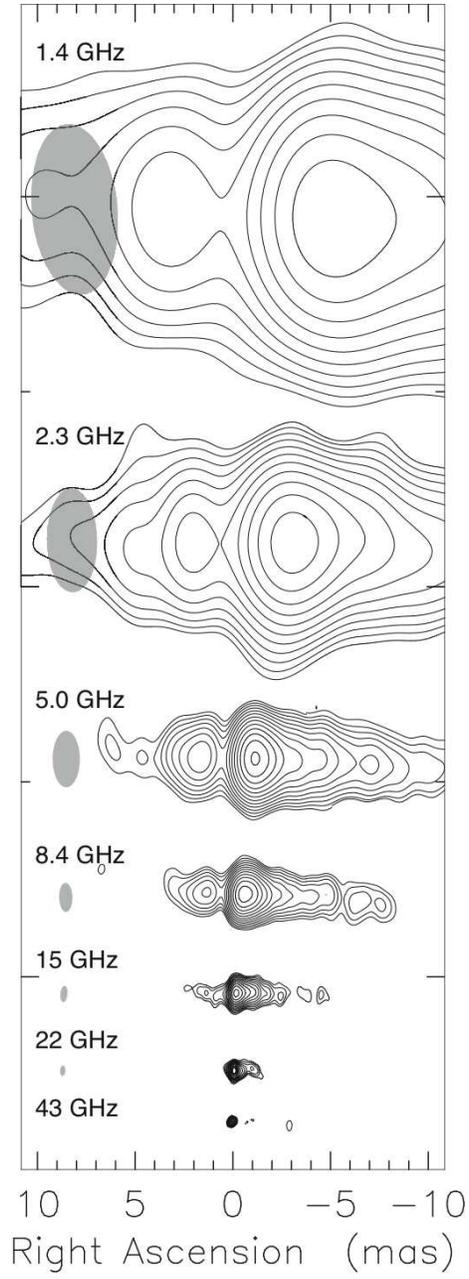}
\caption{
All frequency images, aligned relative to the 43 GHz core (accounting for core shifts).
} 
\label{fig:maps}
\end{figure}

Figure~\ref{fig:all_cln} shows the (self-calibrated) CLEANed images from uniformly weighted visibilities at each frequency; the corrections of relative positions in right ascension, which were determined by phase-referencing (Section~\ref{section:corepositionmeasurement}), have been applied in Figure~\ref{fig:maps}.
The beam sizes and image qualities are listed in Table~\ref{tab:ip}.
Approaching jets and counter jets, extending West and East from the brightest region (core), respectively, appear in the images at $\leq$ 15~GHz clearly.  
The counter jet is not so prominent in the image recorded at 22~GHz, while no obvious jet-like structures are visible in the 43 GHz image. 
However, both approaching and counter jets were clearly detected as model components of {\tt modelfit} in Difmap.  
The lower the frequency, the more extended and symmetric the jets.

\subsection{Core shift measurement}\label{sec:coreshiftmeasurement}
\begin{figure}
\epsscale{.80}
\plotone{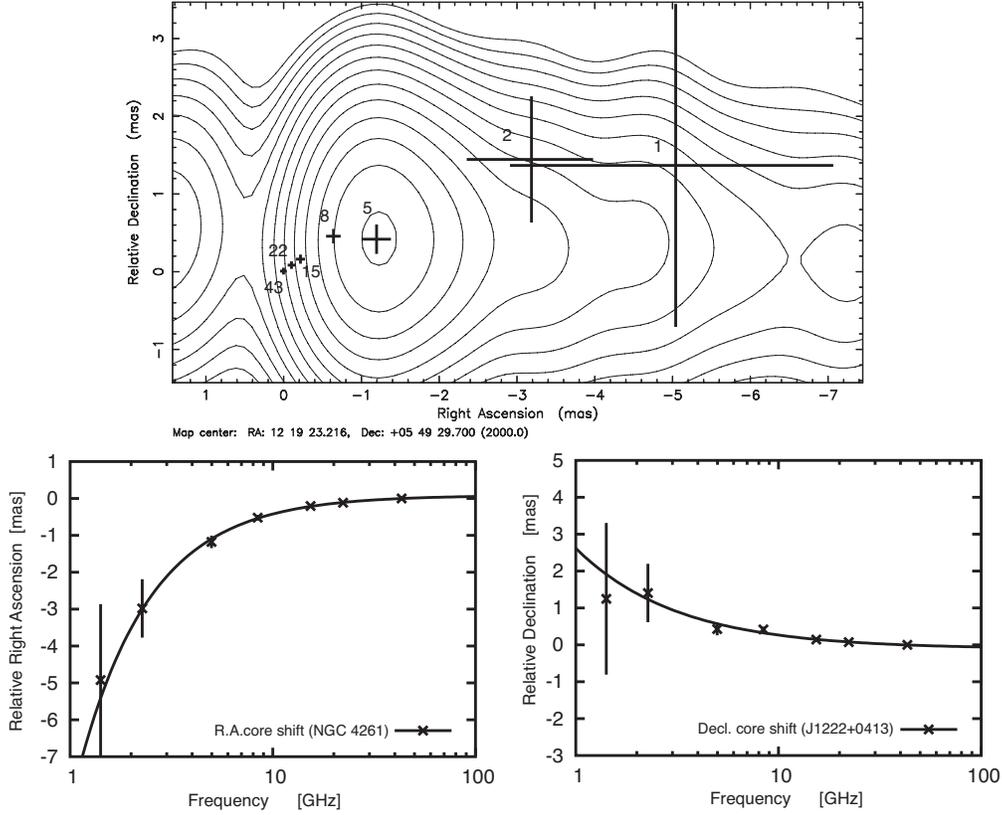}
\caption{
(Top) Crosses denote observed core positions, in approaching jet side at each frequency, relative to the 5~GHz core on the same image.   
The thin contours denote the 5~GHz self-calibrated image.  
The core positions appear to drift northwest as the frequency is lowered. 
Given the jet directions of NGC~4261 and J1222+0413, NGC~4261 is responsible for the core shift in the right ascension direction, while J1222+0413 is responsible for the declination shift. 
(Bottom left) Shift in core position in right ascension as a function of frequency, attributable to NGC~4261. 
The solid line is the least-squares fitted function. 
(Bottom right) Frequency dependence of the declination shift in core position, caused by J1222+0413.
}
\label{fig:cp}
\end{figure}

\begin{figure}
\epsscale{.35}
\plotone{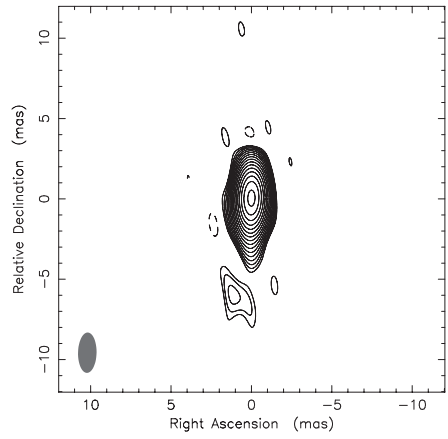}
\caption{
Self-calibrated image of J1222+0413 at 5~GHz. 
The jet extends to the South, almost perfectly perpendicular to the jet of NGC~4261.
}
\label{fig:calib}
\end{figure}

The determined core positions at 1.4--43~GHz are plotted on the 5~GHz self-calibrated image of NGC~4261 in Figure~\ref{fig:cp}.  
The core appears to shift northwestward as the frequency decreases.
The jet of the reference source J1222+0413 is oriented nearly perpendicular to that of \objectname{NGC~4261} (toward the South, whereas the NCG 4261 jet tends to the West; see Figure~\ref{fig:calib}). 
For this reason, the observed core shifts comprise the sum of core shifts of the both sources.   
Therefore, we decomposed the observed core shift into a westward shift for \objectname{NGC~4261} and a northward shift for J1222+0413. 

The east--west structure of \objectname{NGC~4261}'s jet appears slightly fluctuating in North--South direction.  On the approaching side, the jet gradually inclines toward South beyond $\sim4$~mas from the 43~GHz core.  We analyzed the ridge line in the 5-GHz image and measured its deviation to be $-0.24$~mas southward at $-5$~mas where the core at 1~GHz is found.  This southward deviation is used to correct the core position at 1~GHz for the final image to be used for making a spectral index map (Section~\ref{sec:SPIX}).  On the other hand, in the innermost region, the jet shows an S-shaped structure at 15--43~GHz images.  The measured position angle of jet is $-70\degr$ in the 43~GHz image, while $-85\degr$ up to 1.5~mas from the core in the 15~GHz image.  The amplitude of the S-shape is only $\sim50\ \mu$as, which is much smaller than the scale of core shifts in J1222+0413 and NGC~4261.  We do not take into account this North--South deviation for the final results of the core shift and the spectral index maps.   

Figure~\ref{fig:cp} reveals that the core position of \objectname{NGC~4261} shifts at all other observational frequencies relative to the 43 GHz core. 
By least-squares fitting, the core position is found to shift as a power-law function of frequency; that is
\begin{equation}
r(\nu)=\Omega \cdot \nu^{-k}+b,  
\label{equ:cs}
\end{equation}
where $r(\nu)$, $\Omega$, $k$, and $b$ are projected (east--west) distance from the 43 GHz core, free-fitting parameters representing a scale factor, a power low index, and an asymptotic limit at infinite frequency, respectively (solid curves in Figure~\ref{fig:cp}). 
At the approaching jet side, these values are determined as $\Omega=-8.42\pm0.86$, $k=1.22\pm0.06$, and $b=0.082\pm0.016$, respectively, with residuals $\chi^2$/n.d.f = 0.16 (see Table~\ref{tab:cs}). 
This reduced $\chi^2$ is very small because we estimated the core positional errors conservatively.  The core positional errors could be actually smaller than our estimations.  As a result, the free-fitting parameters, $\Omega$, $k$ and $b$, are determined with an uncertainty of 10\%, 5\% and 20\%, respectively.
The value of $k$ suggests a possible excess over the theoretical value of unity predicted for SSA under the equipartition condition and as frequently observed \citep{Lobanov:1998, O'Sullivan:2009, Sokolovsky:2011, Hada:2011}.  
The asymptotic parameter $b$ was determined within a tolerance of 32~$\mu$as ($\approx$ 120 $R_\mathrm{s}$) and separated by 82~$\mu$as ($\approx$ 310 $R_\mathrm{s}$) from the 43~GHz core position.  

We also measured and fitted the core shift at the counter jet side of \objectname{NGC~4261}. 
In this study, the core of the counter jet is defined as the closest east-side component to the core of the approaching jet, the region of maximum intensity yielded by {\tt modelfit}. 
As shown in Figure~\ref{fig:cs}, the distance between the cores in the approaching and counter jets reduces as the frequency increases. 
When the shifting core of the counter jet was fitted to a three-parameter model, we obtained $\Omega=-8.34\pm6.40$, $k=0.16\pm0.23$, and $b=-4.34\pm7.38$, respectively.
However, these errors have little contribution for confining the position of the central engine.  This suggests that a simple power-law is insufficient to represent the profile of counter-jet side.  
Hence, we applied two alternative model fitting procedures imposing further assumptions, which will be discussed in Section \ref{sec:cs}.

\subsection{Spectral index maps}\label{sec:SPIX}
Spectral index ($\alpha$, $S \propto \nu^{\alpha}$) maps were generated from final (self-calibrated and CLEANed) images between two adjacent frequencies using the AIPS task {\tt COMB}. 
To ensure proper construction of spectral index maps, the images between the adjacent frequencies must spatially align, and their angular resolutions must match. 
The map coordinates at each frequency were aligned using the core shift relative to the 43~GHz core in right ascension and the jet axis of \objectname{NGC~4261} in declination. 
The images at adjacent frequencies were convolved by rendering both beams equivalent to the beam at the lower frequency. 
The intensity below 3$\sigma$ was clipped out before calculating the spectral index.

\begin{figure}
\epsscale{0.9}
\plotone{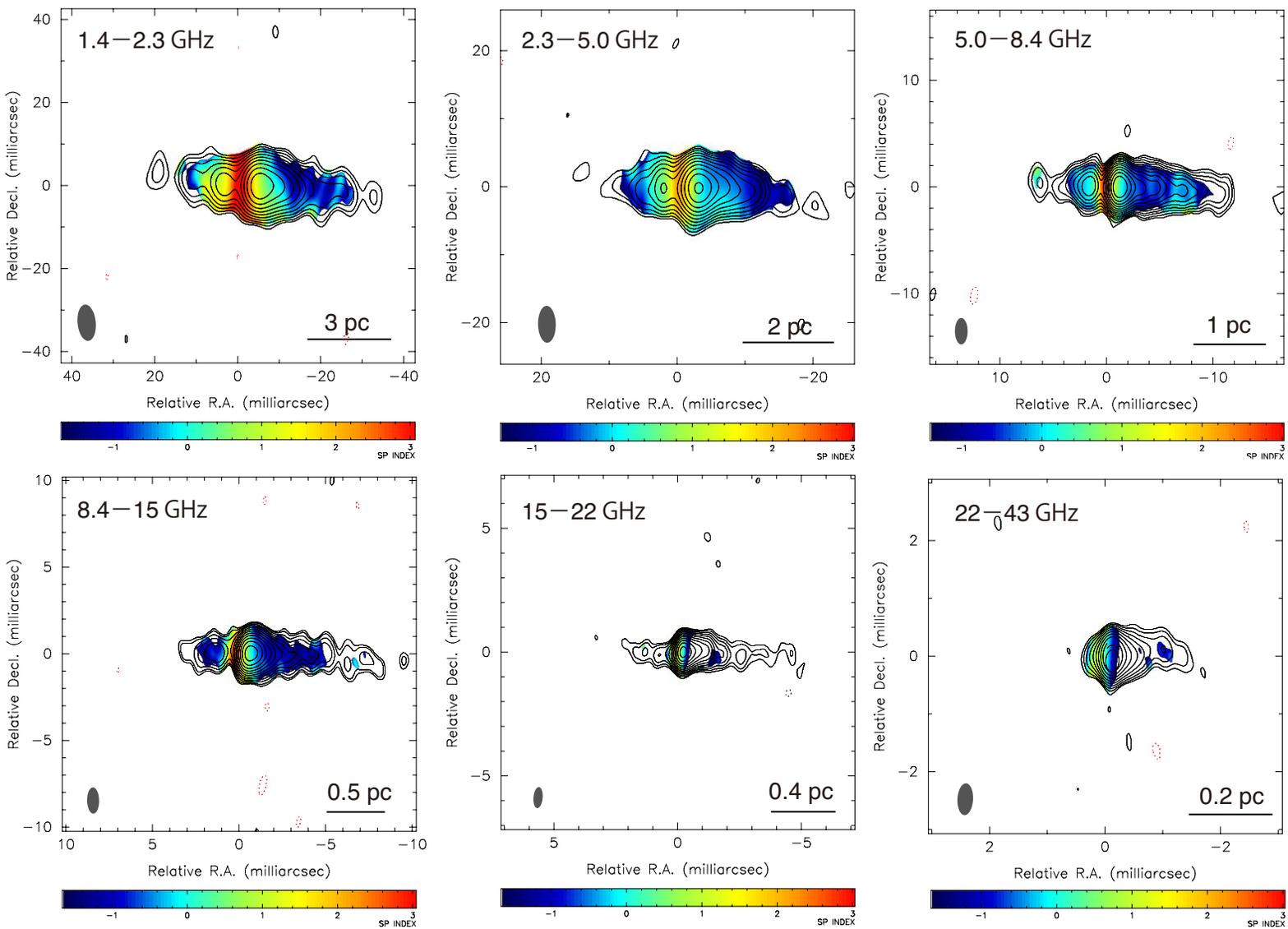}
\caption{Spectral index maps of all adjacent frequencies.  Overlaid contours are total intensity at the lower frequency.   
}
\label{fig:all_spx}
\end{figure}

All of the spectral index maps exhibit common features, in particular, the spectral index peaks in the regions between the cores of the approaching and counter jets. 
As the jet extends further from the core, the spectral index becomes smaller. 
Figure~\ref{fig:all_spx} displays the spectral index maps of the 1--2~GHz and 5--8~GHz pairs. 
Some regions show $\alpha> 2.5$, which is the theoretical limit of SSA. 
This result is consistent with other observations in which the circumnuclear structure favored FFA over SSA \citep{Jones:2001}. The spectral index gradients are clearly parallel to the jet axis for all frequency pairs.
The restored beams in each map are too large to decompose the absorption structures across the jet.  
However, a transverse gradient marginally appear at 1--2 GHz and 2--5 GHz spectral index maps.  We cannot rule out that the combination of the North--South fluctuation in the jet structure (Section~\ref{sec:coreshiftmeasurement}) and remaining position errors in East--West direction causes the apparent transverse gradient.

\begin{figure}
\epsscale{0.6}
\plotone{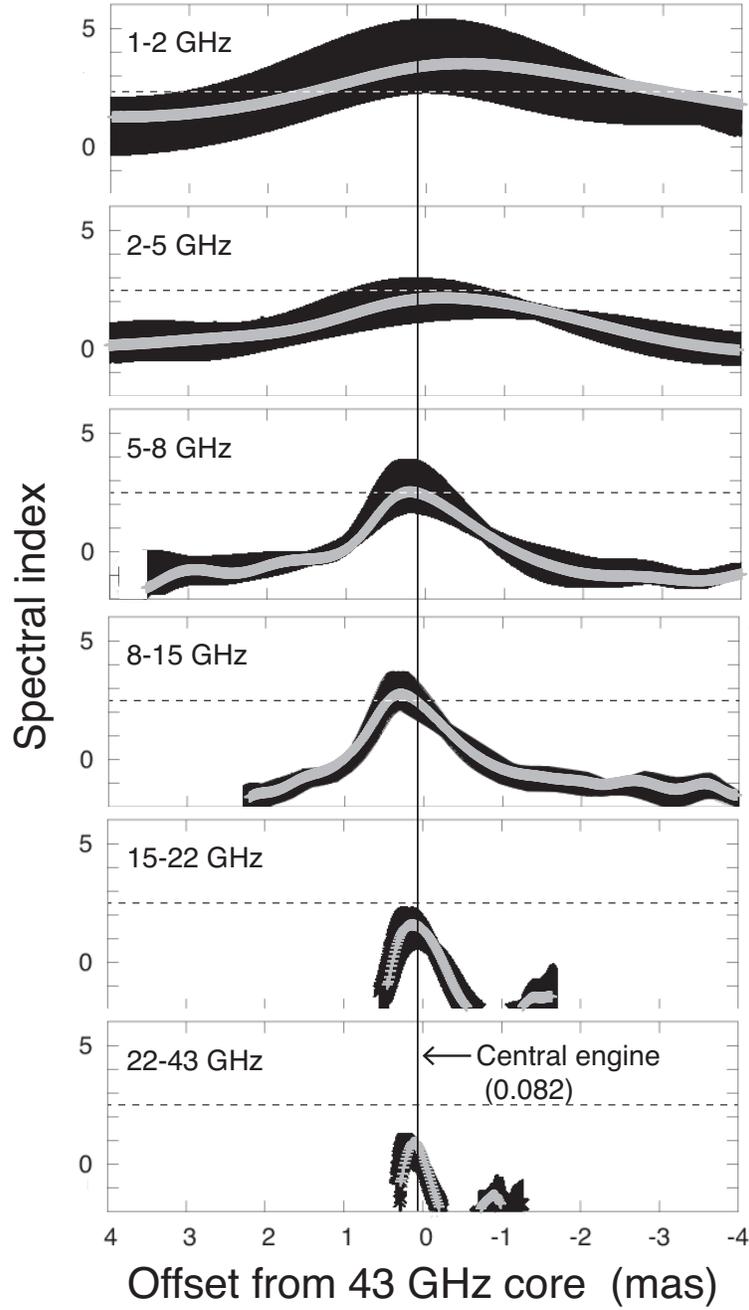}
\caption{
Spectral index profiles along jet axis at each pair of adjacent frequencies.   
Black regions represent estimated spectral-index errors, consisting of the accuracy of core position, the uncertainty of flux scaling, and image noises (see, Section~\ref{sec:SPIX}).
Gray narrow lines represent median values. 
}
\label{fig:slice}
\end{figure}

Figure~\ref{fig:slice} shows slice profiles of spectral index along the jet axis (from the East to the West). 
We estimated error components in the spectral index. 
We tentatively displaced one image eastward and westward with respect to the other image by our conservative positional error, in order to evaluate a spectral index error due to the positional error in overlaying two images.         
We also used the spectral error map, simultaneously produced during {\tt COMB} processing, and assumed a conservative flux calibration error of 5\% at $\nu \leq 15$~GHz and 10\% at $\nu \geq 22$~GHz.  
The total error represented by an envelope is the square root of the summed squares of these error components.  

At 8--15~GHz and lower frequency pairs, a region of $\alpha> 2.5$ appears in the slice profiles, suggesting an FFA mechanism, although the error is too high for a conclusive assessment. 
Another feature is that the curve at $\nu \geq$ 5~GHz peaks at the counter jet side. 
The peaks are located at the same positions, regardless of the error-induced displacement of the core position. 
At $\nu \leq$ 5~GHz, the peaks are located around the origin.

%\subsection{Formalism} \label{bozomath}

%% The equation environment wil produce a numbered display equation.

%% The \notetoeditor{TEXT} command allows the author to communicate
%% information to the copy editor.  This information will appear as a
%% footnote on the printed copy for the manuscript style file.  Nothing will
%% appear on the printed copy if the preprint or
%% preprint2 style files are used.

%% The eqnarray environment produces multi-line display math. The end of
%% each line is marked with a \\. Lines will be numbered unless the \\
%% is preceded by a \nonumber command.
%% Alignment points are marked by ampersands (&). There should be two
%% ampersands (&) per line.

%% Putting eqnarrays or equations inside the mathletters environment groups
%% the enclosed equations by letter. For instance, the eqnarray below, instead
%% of being numbered, say, (4) and (5), would be numbered (4a) and (4b).
%% LaTeX the paper and look at the output to see the results.

%% This section contains more display math exaFmples, including unnumbered
%% equations (displaymath environment). The last paragraph includes some
%% examples of in-line math featuring a couple of the AASTeX symbol macros.

\section{Discussion}
%%%%%%%%%%%%%%%%%%%%%%%%%%%%
%4.1
\subsection{Position of the black hole in NGC~4261}\label{sec:posi_bh}
First, we consider the relationships between the black hole positions and limiting core positions. 
\citet{Hada:2011} showed that the central engine in \objectname{M87} is located $23~R_\mathrm{s}$ distant from the observed core at 43~GHz in their analyses of core shift measurements. 
This study assumed that the asymptotic line, representing the origin of a jet, spatially coincides with the central engine. 
However, the core location in \objectname{BL Lac}, indicated as the standing shock point \citep{Marscher:2008}, is actually $10^4$--$10^6~R_\mathrm{s}$ from the black hole, as determined from $\gamma$-ray monitoring observations \citep{Albert:2007}. 
On the other hand, a core shift was discovered downstream of the \objectname{BL Lac} jet. 
Core shift measurements of \objectname{BL Lac} and other blazar sources revealed that the base of the jet is located $10^{4}~R_\mathrm{s}$ from the 43 GHz core position \citep{O'Sullivan:2009}. 
Recently, core shift data have accumulated at an accelerated pace for blazars \citep{Sokolovsky:2011, Pushkarev:2012}. 
However, the above-cited examples highlight the difficulty of proving assumptions from one-sided jets.

\begin{figure}
\epsscale{.50}
\plotone{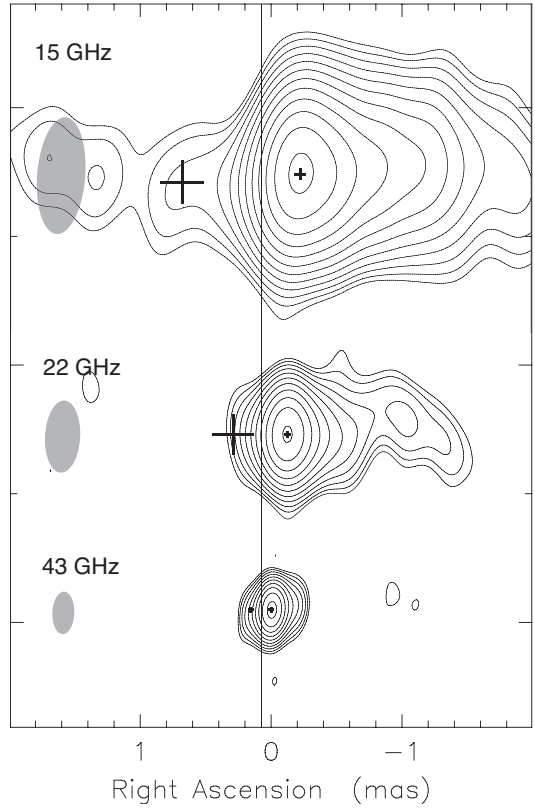}
\caption{
Enlarged image from a part of Figure~\ref{fig:maps} for $\nu \ge$ 15~GHz within $\pm$ 2 mas. 
The cross marks denote determined positions of innermost cores in each jet side. 
The line at R.A. $= 0.082\pm0.016$ mas represents the determined position of the central engine. 
}
\label{fig:large}
\end{figure}

To overcome these limitations, we located the central engine in \objectname{NGC~4261} by investigating a double-sided system and measuring the core positions of approaching jet and counter jet. 
As shown in Figure~\ref{fig:large}, the core (most upstream component) at the counter jet side resides at $0.17\pm0.03$~mas ($\approx 640\pm110~R_\mathrm{s}$) from the approaching-jet core at 43 GHz. 
The central engine and black hole must be situated at most 750~$R_\mathrm{s}$ distant from this core to ensure its location between the cores of both jets. 
Moreover, because the origin of the counter jet coincides with that of the approaching jet (as discussed later), the assumption that the origin of a jet spatially coincides with the central engine within the error tolerance is reasonable for \objectname{NGC~4261}. The center of activities were discussed using cross-identificatiton of the Gaussian component models in multifrequency maps in other two-sided jet systems (e.g., \objectname{Cygnus A}; \citealt{Krichbaum:1998} and \objectname{NGC~1052}; \citealt{Kadler:2004}).  
For \objectname{Cygnus A}, a relatively distant source, the core shift in the counter jet has not been measured. \citet{Kadler:2004} determined the position of the central engine in NGC~1052 with an uncertainty of $\sim0.03$~pc, corresponding to $\sim2000\ R_\mathrm{s}$ assuming $\log{M_\mathrm{BH}/M_{\odot}}=8.19$ \citep{Woo:2002}, from the both sides of jets. Our result for NGC~4261 is of higher accuracy and includes coordinates relative to 1222+0413 rather than relative to optically thin components in the source itself.

%\textcolor{blue}{m9) you should discuss any measurements in the literature on measurements of the center of activity in other 2-sided jet systems (Cygnus A?, NGC1052?,...)}

\subsection{Core position at 43 GHz}\label{sec:core_43}
The central engine of \objectname{NGC~4261} is located at  $82\pm16~\mu$as, corresponding to $310\pm 60~R_\mathrm{s}$ from the 43 GHz core. 
The core--central engine separation differs by orders of magnitude among \objectname{NGC~4261}, \objectname{M87}, and \objectname{BL Lac}, being larger and smaller in \objectname{M87} and \objectname{BL Lac}, respectively, than in \objectname{NCG~4261}. 
The separation caused by a core shift is correlated with the strength of SSA, and depends on jet characteristics such as luminosity, velocity, structure, radial profiles of density and magnetic field, the geometric relationship to our line of sight, velocity gradients, and/or their time variation. 
According to \citet{Lobanov:1998}, if a jet is continuous, uniform, stationary, and travels with constant velocity, and under equipartition condition, the (projected) core shift from the jet origin is given by %equipart

%According to Lobanov~1998, assuming that the jet has the continuous, uniform, stationary state and  a constant velocity, addition to the equipartition between jet particle and magnetic field energy density, the core shift from the origin of a jet can be described as
%
%
\begin{equation}
\Delta r_\mathrm{mas} \approx \frac{C_\mathrm{r} (1+z)}{D_L \gamma^{2} \phi_o \nu_1} \left[\frac{L_\mathrm{syn} \sin \theta}{\beta(1-\beta \cos \theta)\Theta} \right]^{2/3},
\label{equ:lob}
\end{equation}
where $C_\mathrm{r} = 4.56\times10^{-12}$, and $\gamma \equiv 1/\sqrt{1-\beta^2}$, $\beta \equiv v/c$, $\phi_o$, and $\theta$ denote the Lorentz factor, the intrinsic jet velocity, jet opening angle from an observer (intrinsic opening angle $\phi=\phi_o/\sin{\theta}$), and viewing angle, respectively. 
$D_\mathrm{L}$ and $L_\mathrm{syn}$ are the luminosity distance to the source (parsec) and synchrotron luminosity (erg s$^{-1}$). 
In this case, $\Theta$ depends on the jet structure, whether conical or parabolic, and does not usually exceed ln~3 for parsec-scale jets. 
The frequency $\nu$ is given in Hz.  
In the present study, the separation between the central engine and core position is determined at 43 GHz.  
Using Equation~(\ref{equ:lob}), we now compare the separation in \objectname{NGC~4261} to those estimated in \objectname{M87} and \objectname{BL Lac}.  
In place of $L_\mathrm{syn}$, we use the intrinsic luminosity at millimeter wavelengths, $L_\mathrm{m}$. 
In order to remove beaming effect from the observed luminosity $L^{\dagger}_m$, we calculate the beaming factor $D^{3-\alpha}$, where $D=[\gamma (1-\beta \cos \theta)]^{-1}$ is the Doppler factor, using the viewing angle and intrinsic velocity for each source.  
Using $\theta = 63^\circ$, $\beta=0.46$ for \objectname{NGC~4261} \citep{Piner:2001} and $L^{\dagger}_\mathrm{m} = 3.9\times10^{40}$~erg s$^{-1}$ \citep{Geldzahler:1981}, we estimated a beaming factor of 1.5 and $L_\mathrm{m} = 2.4\times10^{40}$~erg s$^{-1}$, assuming $\alpha =-0.7$.
The parameters of \objectname{NGC~4261}, \objectname{M87}, and \objectname{BL Lac} are provided in Table~\ref{tab:core43}.

 As a rough discussion, assuming that the jet opening angle is nearly constant\footnote{Actually, the jet profile of \objectname{NGC~4261} is nearly conical, which will be reported in an another paper.} under the Equation~(\ref{equ:lob}), the ratio of the core--central engine (deprojected) separation to that of \objectname{NGC~4261} in $R_\mathrm{s}$ unit is given by
\begin{eqnarray}
%\begin{equation}
%
\Delta r_\mathrm{ratio} &\approx&  1.92 \times (1+z)^{-1} \left(\frac{(1-\beta^2)^{3/2} \sin\theta}{\beta(1-\beta\cos\theta)}\right)^{2/3} \nonumber\\
&&\hspace{1.5cm} \times \left(\frac{M_\mathrm{BH}}{10^8 M_\odot}\right)^{-1} \left(\frac{L_\mathrm{m}}{10^{40}~\mathrm{erg/s}}\right)^{2/3}.
\label{equ:ratio}
%
%\end{equation}
\end{eqnarray}

For M87 case, the viewing angle of the jet is likely $15^\circ$--$25^\circ$\citep{Acciari:2009} and apparent jet velocities of $\beta_\mathrm{app} = 0.25$--$0.40$ have been reported for the innermost region at 43~GHz \citep{Ly:2007}.  
Under these situation, the intrinsic velocity would be $\beta = 0.38$--$0.62$. 
Assuming $\theta = 25^\circ$, $\beta = 0.38$ and $L^{\dagger}_\mathrm{m} = 5.2\times10^{40}$~erg s$^{-1}$ \citep{Davies:2009}, we obtain $L_\mathrm{m} \approx 1.5\times10^{40}$~erg s$^{-1}$.
Adopting a black hole mass of $6\times10^9 M_\odot$ \citep{Gebhardt:2009}, we obtain $\Delta r_\mathrm{ratio} = 0.05$.
For $\theta = 15^\circ$ and $\beta = 0.62$, we obtain $L_\mathrm{m} \approx 0.43\times10^{40}$~erg s$^{-1}$ and $\Delta r_\mathrm{ratio} = 0.01$.
On the other hand, based on the core shift measurements, the 43-GHz core--central engine distance ratio of $23 R_\mathrm{s}$ (\objectname{M87}) to $310 R_\mathrm{s}$ (\objectname{NGC~4261}) is 0.07. 
The difference between the two results is not so large.  
The discrepancy may originate in uncertainties of the black hole mass or the intrinsic luminosity.
Another possibility is the difference of  opening angles between different sources.
From Equation~(\ref{equ:lob}), the separation is inversely proportional to the opening angle.
If the opening angle of \objectname{M87} is widely larger \citep[cf.][]{Junor:1999} than that of \objectname{NGC~4261} it can explain the difference of the separation between them.

By contrast, the core shift in \objectname{BL Lac} substantially differs from that of \objectname{NGC~4261}. 
% The $\Delta r_\mathrm{ratio}$ of \objectname{BL Lac} is 30--3000. 
The estimated jet velocity and inclination are $\beta = 0.99$ and $\theta = 6^\circ$--$10^\circ$, respectively \citep {Jorstad:2005}. 
Such conditions should enhance the relativistic beaming effect, and indicate a beaming factor of $3.6\times10^3$.   
Since the observed luminosity was reported to be $3.6\times10^{43}$~erg s$^{-1}$, we estimate an intrinsic luminosity as $L_\mathrm{m}\approx 1\times10^{40}$~erg s$^{-1}$.
The mass of the black hole associated with \objectname{BL Lac} is estimated to be $1.7\times10^8 M_\odot$ \citep{Woo:2002}.  
Substituting these value into Equation (\ref{equ:ratio}), we obtain $\Delta r_\mathrm{ratio} \approx 0.08$ as an expected core shift, showing a large discrepancy from the result from $\gamma$-ray monitoring observations ($\sim30$--3000)  \citep{Albert:2007,Marscher:2008}.  
To explain the large discrepancy, a much larger intrinsic luminosity of $L_\mathrm{m} \ge 7\times10^{43}$ erg s$^{-1}$ or a much smaller opening angle $\phi_o \le 0.08$~deg is necessary.

However, such a wide difference in intrinsic luminosity between \objectname{NGC~4261} and \objectname{BL Lac} is quite unlikely if both FR-I radio galaxies and \objectname{BL Lac} objects are regarded as a common AGN population intrinsically \citep{Urry:1991}. 
More likely, the appearance of FR-I radio galaxies and \objectname{BL Lac} objects differ because of the beaming effect imposed by jet inclination.  
The jet of \objectname{M87} has a limb-brightened structure, which is probably associated with a fast spinal flow surrounded by a slow layer (sheath) \citep{Ghisellini:2005}. 
If the jet axis inclines significantly from the line of sight, for such as \objectname{M87} and \objectname{NGC~4261} as FR-Is, sheath emission dominates because spinal emission is deboosted by the relativistic effect. %de-boost 
On the other hand, if the jet is viewed from nearly pole-on, for such as \objectname{BL Lac}, spine emission from a small opening angle dominates. 
This phenomenon can explain the varying distance between the central engine and the 43~GHz core between FR-I radio galaxies and \objectname{BL Lac} objects.  

%%%%%%%%%%%%%%%%%%%%%%%%%%%%
%4.3
\subsection{What cause the core shift}\label{sec:cs}
The physical causes of the core shift are considered as SSA and/or FFA. 
The core shift measurement for a two-sided system allow us to utilize a symmetry assumption  for understanding the jet structure and surrounding circumstance, and offers a critical advantage over studies for one-sided systems.   
Only the counter jet itself could not well describe its core shift profile (Section~\ref{sec:coreshiftmeasurement}).
In the following subsections we establish the model that best explains the core shift of \objectname{NGC~4261}, focusing on the results of fitting parameters in Equation (\ref{equ:cs}).   

%The symmetrically emanated jets guarantee that the core shifts in both sides can be described with common parameters except for one in terms of inclination opposed each other.  
%
%In the assumption that the jet properties of both sides are intrinsically identical, the core shift parameter $k$ should be the same values for the both sides.  Therefore, we can predict an intrinsic core shift for the counter side on the basis of knowledge of the approaching side, and then, the physical process causing the observed core shift is expected to be judged by comparing the two sides.    
%
%Although the core shift could not be closely fitted at the counter jet side (Section~\ref{}),  
%
%M4) The fitting parameter $k$ does not suggest a common value between the two sides (Section \ref{sec:cs}).  
%Here, 
%The parameters at counter jet side can be estimated from those of the approaching side in each model.

\subsubsection{SSA in Jets}\label{sec:SSAinjet}
First, we assume a pure SSA model. 
The physical conditions in the jet reflect the index of the core shift profile in Equation (\ref{equ:cs}).
In this model, $k_s$ (representing $k$ for SSA case) is given by
\begin{equation}
k_\mathrm{s} = \frac{5-2\alpha}{(3-2\alpha)m+2n-2} ,
\label{equ:k_ssa}
\end{equation}
where $\alpha$, $m$ and $n$ denote the spectral index at the optically-thin regime, the profiles of the magnetic field $B$ (where $B \propto r^{-m}$), and the electron density $N$ (where $N \propto r^{-n}$), respectively \citep{Blandford:1979}. 
Assuming $m=1$ and $n=2$ as the profile indices, corresponding to the stabilization of jet energy in equipartition, $k_\mathrm{s}$ invariably becomes unity regardless of the spectral index value.
However, the fitting parameter at the approaching jet side is $k_\mathrm{s}=1.22 \pm 0.06$, which possibly exceeds that from previous studies for other sources ($k_\mathrm{s}\approx1$; i.e., \objectname{M87}; \citealt{Hada:2011}, \objectname{BL Lac}; \citealt{O'Sullivan:2009}, and \objectname{3C~345}; \citealt{Lobanov:1998}), indicates that $m$ and $n$ in Equation (\ref{equ:k_ssa}) can take different values in \objectname{NGC~4261}.

We next predict the core shift profile at the counter jet side assuming the SSA model. 
If the jets are intrinsically symmetric, $k_\mathrm{s}$ and $b$ of the counter and approaching jets side should be identical. 
On the other hand, comparing Equations (\ref{equ:cs}) and (\ref{equ:lob}), we find that $\Omega$ depends on geometrical and intrinsic parameters of a jet. 
The parameter $\Omega$ differs between the approaching and counter jets only through the inclination, $\theta$. 
Therefore, $\Omega$ of the counter and approaching sides, denoted $\Omega_{\mathrm{CJ}}$ and  $\Omega_{\mathrm{AJ}}$, respectively, must satisfy  $\Omega_{\mathrm{AJ}} >  \Omega_{\mathrm{CJ}}$. 
In other words, the position of the core at the same frequency is farther from the black hole in the approaching jet than in the counter jet. 
The parameter $\Omega_\mathrm{CJ}$ should be connected with $\Omega_\mathrm{AJ}$ by a difference of 180\degr\ in terms of viewing angle.  
The curve predicted by the pure SSA model (dashed line in Figure~\ref{fig:cs}) can explain the core position at both low frequency ($\leq$ 2~GHz) and high frequency ($\geq$ 22~GHz). 
On the other hand, the observed positions of the core at 5, 8, and 15~GHz are significantly larger than those predicted by this model, implying that the SSA model alone lacks sufficient absorption to explain the core shift in these regions.  

\subsubsection{Spherical FFA plasma}\label{Sec:dis_ffa}
Another alternative is the FFA model. 
The spectral index maps indicate that FFA plasma exists at least in regions near the putative black hole. 
For simplicity, and from the parameters of the core shift measurement, we assume a spherical FFA plasma (see, Appendix~\ref{Sec:App_ffa}). 
In this case, $k$ depends on the structure of the spherical FFA plasma. 
When the FFA opacity $\tau_\mathrm{f}$ becomes unity, $k_f$ (representing $k$ in Equation (\ref{equ:cs}) for the FFA model) is given by

\begin{equation}
k_\mathrm{f} = \frac{2.1}{2s-\frac{3}{2}t-1},
\label{eq:kf}
\end{equation}
where $s$ and $t$ represent the profiles of the electron density $n_\mathrm{e}$ and temperature $T_e$ ($n_e \propto r^{-s}$, $T_e \propto r^{-t}$) in the FFA absorber, respectively. 
These parameters can be determined from the opacity profile; the denominator in the right side of Equation~(\ref{eq:kf}) represents the index value of the opacity profile ($\tau_\mathrm{f}=2s-\frac{3}{2}t-1$). 
We obtain the radial profile of FFA opacity at the approaching jet side by fitting to a power-low function; $\tau_\mathrm{f} \propto r^{-1.6}$ (Appendix~\ref{Sec:App_ffa}).
Substituting this value into Equation~(\ref{eq:kf}), we obtain $k_\mathrm{f} = 1.3$.
It is consistent with the result of the core shift measurements ($k_\mathrm{f} = 1.22\pm0.06$), in spite of independent measurements of the core shift and the opacity.  
Theoretical speculations have indicated that $k_\mathrm{f}$ exceeds unity \citep{Lobanov:1998, Kadler:2004}. 
Thus, the core shift at approaching also jet side can also be sufficiently interpreted by the spherical FFA model.

However, the spherical FFA is inconsistent with the radial profile of FFA opacity on the counter jet side. 
The parameter $k_\mathrm{f}$ should be identical in both the approaching and counter sides if the FFA absorber is symmetric.  $\Omega$ reflects the path length of the absorber; that is, the geometrical structure of the absorber (for details, see Appendix). 
If the absorber covers the counter jet side, $\Omega$ must satisfy $\Omega_{\mathrm{AJ}}<\Omega_{\mathrm{CJ}}$ because the path length of the counter jet side is longer than that of the approaching one.  
The core-shift curve predicted by the spherical FFA plasma model (dot--dashed line in Figure~\ref{fig:cs}) meets some of core positions at $\geq$ 5~GHz on the counter jet side. 
However, the core shifts at 1 and 2 GHz are inconsistent because their shifts are significantly smaller than those predicted by this model.  

Therefore, neither of these simple models can explain the core shifts in this two-sided system. 

\subsubsection{SSA in Jet + FFA by the disk}\label{sec:SSA+FFA}
A possible solution is to combine the SSA core shift with an additional external absorber, which compensates for the inadequate core shift. 
Indeed, our results show that the pure SSA model fitted fairly well on both approaching and counter sides at all frequency, except for the counter-jet components at 5, 8, 15~GHz.   
When the observed core shifts in the counter side is fitted without the cores at these frequencies, the best-fit values of $k = 1.15\pm0.06$ and $b = 0.08\pm0.02$ (representing the index and the asymptotic limit in this condition) approximate those of the approaching side, as expected from SSA theory (Table~\ref{tab:cs}).  
This suggests the presence of an absorber in a limited region. 
Thus, it appears that the cores at 5, 8, and 15 GHz are additionally affected by FFA. 
FFA opacity $\tau_\mathrm{f}$ at the positions of 2--22~GHz cores in the counter jet is determined by spectral analysis of multi-frequency images (Appendix~\ref{sec:analysisFFA}) and shown in Figure~\ref{fig:tau}.  On the 5--15 GHz core, $\tau_\mathrm{f}$ is well represented by a power-law profile.  On the other hand, $\tau_\mathrm{f}$ at other frequencies, where SSA alone can describe the core shift profile, should be $\tau_\mathrm{f} < 1$.  

Such a limited range of FFA regions is consistent with a gap found in the intensity profile in the counter jet side in previous studies \citep[e.g.,][]{Jones:2001} and an obscuring disk proposed by them.   
Thus, the geometrically thin disk predicted is a candidate to explain the core shift of the counter side.
 %%%%%%%%%%%%%%%%%%%%%%%%%%%%
%4.4

\subsection{Disk structure of NGC 4261}\label{sec:diskstructure}
We consider the structure of a disk to explain the core shift and the profile of $\tau_\mathrm{f}$ at the counter jet side.

\subsubsection{Disk Model}
Figure~\ref{fig:disk} shows the geometric relationships between the jet and disk, assuming $\theta=63^\circ$ and the rotation axis of the disk is aligned with the axis of pc-scale jet. 
As discussed in the combined model, if the disk plasma cause the core shift at 5--15~GHz, the density and temperature of the disk require the radial gradient ($n_\mathrm{e} = n_1 x^{-s}$, $T_\mathrm{e} = T_1 x^{-t}$, where $x$ is the radius of the disk in $R_\mathrm{s}$ unit, $r=xR_\mathrm{s}$). 
In addition, the thickness of the disk and the path length through the disk to observers are also described as $H = H_1 x^{-h}$ and $L= H \sec \theta$.
By substituting $n_\mathrm{e}$, $T_\mathrm{e}$ and $L$ into Equation (\ref{eq:tau_f}), we obtain $\tau_\mathrm{f}$ by the disk (hereafter $\tau_\mathrm{disk}$); that is  

\begin{equation}
\tau_\mathrm{disk} = 0.46 n_1^2 T_1^{-3/2} H_1 \sec \theta x^{-(2s-3t/2-h)}.
\label{equ:tau_disk}
\end{equation}

\begin{figure}
\epsscale{0.4}
\plotone{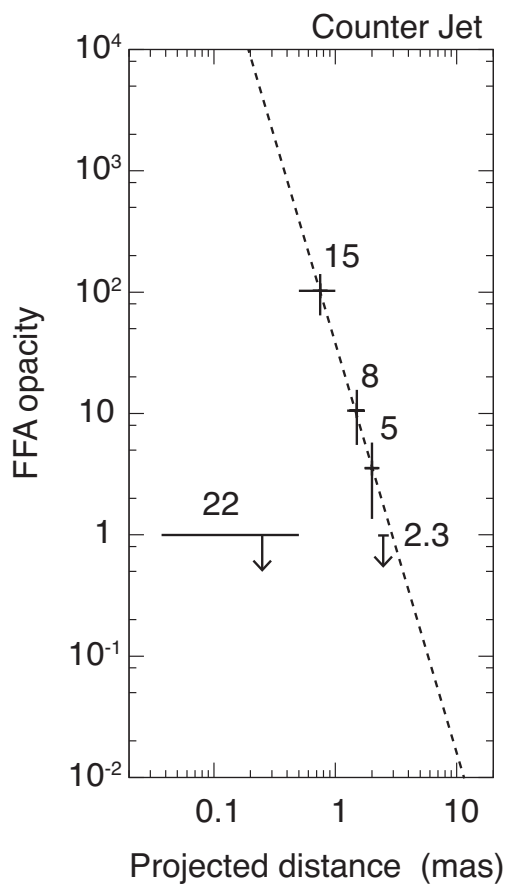}
\caption{
FFA opacity at the locations of cores in the counter jet side, assuming that FFA is responsible for observed deviations of core positions at 5, 8, and 15~GHz (Section~\ref{sec:SSA+FFA}) and the spectral profile along counter jet is caused by an FFA process.      
The upper limits at the low frequency (2~GHz) and the high frequency (22~GHz) are because of no significant deviation from the SSA core shift model.    
The dashed line denotes a possible radial profile of FFA opacity as a power-low function by least-square fitting with 5--15~GHz data.  
}
\label{fig:tau} 
\end{figure}

In the Figure~\ref{fig:tau}, the least-squares fitting at at 5--15~GHz data reveals the profile of disk opacity to the power-law function of the disk radius; that is
$\tau_\mathrm{f} \propto r^{-3}$, then we obtain 
\begin{equation}
2s-\frac{3}{2}t-h = 3 .
\label{equ:tau_ind}
\end{equation}

\subsubsection{Core shifts at $\nu <$ 2~GHz and $\nu >$ 22~GHz do not require disk absorption}
To explain the small opacity at low frequency ($\leq$ 2 GHz), the low density, the short path length or high temperature are required because the FFA effects depend on frequency ($\tau \propto \nu^{-2.1}$, see Appendix~\ref{sec:analysisFFA}).
In these requirements, the condition of the path length and the temperature are physically unrealistic because the low frequency core positions correspond to the outer radii of the disk.
On the other hand, it is realistic that the density is sufficiently small and/or the temperature of the plasma is too low to be ionized at the outer radii.  

We consider the condition that the temperature at the radius corresponding to the 5~GHz core position and at the 2~GHz are higher and lower than $10^4$~K, respectively.
Assuming that $t=3/4$ predicted by standard disk model at the outer region \citep{Shakura:1973} and $10^4$~K at 5 GHz, we obtain
\begin{equation}
T_\mathrm{e}= 1\times10^7 x^{-3/4}\ (\mathrm{K}).
\label{equ:T}
\end{equation}
In addition, if $h=0$ (wherever the path length is constant) and $H_1 = 0.01$~pc, for simplicity, we obtain $s= 2$ from Equation (\ref{equ:tau_ind}).
It is very close to $s=15/8$ predicted by standard disk model.
Inserting these value and $\tau_\mathrm{disk}$ to Equation (\ref{equ:tau_disk}), we obtain 
\begin{equation}
n_\mathrm{e}= 8\times10^{12} x^{-2}~~(\mathrm{cm}^{-3}).
\label{equ:N}
\end{equation}
The density and temperature profile in this condition is shown in Figure~\ref{fig:disk}.

We note that the assumptions of $h$, and $H_1$ may be unsupported and thus other values are acceptable.
Substituting $h = -9/8$ (predicted by standard disk model) into Equation~(\ref{equ:tau_ind}) and $H_1 = 2\times10^{-6}$~pc (corresponding to $L = 0.01$ pc at the 5~GHz core), we obtain $s = 2.6$ and $n_1 = 2\times10^{15}$~cm$^{-3}$.
Although the index $s$ increases the gap to the value predicted by the model, $n_\mathrm{e}$ at $x = 10^3$--$10^5$ is almost the same within a few times factor comparing to the case of $h=0$. 
In the case of $H_1$, $n_\mathrm{e} \propto H_1^{-0.5}$ is obtained from Equation (\ref{equ:tau_disk}), indicating the thinner disk geometrically requires the higher density (i.e., $n_1 = 8\times10^{13}$~cm$^{-3}$ at $H_1 = 10^{-4}$~pc).
However, if the disk thickness was much larger (i.e., $H_1 > 1$~pc), the disk would obscure the approaching jet as well as the counter jet.
Thus, $H_1 = 0.01$ pc provide a reasonable upper limit, indicating the lower limit of $n_\mathrm{e}$.

On the other hand, this density profile cannot explain the small opacity at the high frequency ($\geq$ 22 GHz, $x= 1.5\times10^3$). 
The interpretation under a simple pow-low profile requires a large $t$ such as $t \geq 3$, which leads to
 a sufficiently high temperature of plasma to reduce FFA effect at the inner region (22~GHz).  However, this is impractical because the disk temperature at 3~$R_\mathrm{s}$ should be too high ($10^{14}$~K).  
%  and the temperature is too low to ionize gas ($T < 10^4$ K) at outer region (2~GHz), where significant FFA is observed.      
%
The weak effects of FFA at the 22~GHz core indicate that the disk structure changes dramatically at a characteristic radius.    

The RIAF model is suitable for describing this situation of the low density at inner radii of an accretion.  
In fact, X-ray observations suggest that a low luminosity and flux and spectral variations of the nuclear X-ray emission of NGC~4261 can be caused by an accretion disk with a low accretion rate with a low radiative efficiency such as RIAF \citep{Gliozzi:2003}.  
Therefore, we propose the accretion flow of NGC~4261 has a two-component structure; one is an inner, hot, optically thin, RIAF and the other is an outer, truncated, cool, optically thick accretion disk. 
This structure is considered to be not expedient for NGC~4261 but common for LLAGNs based on broad spectral energy distribution~(SED) and X-ray spectral studies \citep{Nemmen:2006,Nandra:2007}.

\subsubsection{Comparison with theoretical models for transition radius in accretion flow/disk}
According \citet{Liu:1999}, the transition radius, $x_c$, in an accretion disk is given by 
\begin{equation}
x_c=18.3 m_b^{0.17/1.17}\dot{m}_b ^{-1/1.17}
\label{equ:xc}
\end{equation}
where $m_b$ and $\dot{m}_b$ denote the black hole mass in units of solar masses, $M_\mathrm{BH} = m_b M_\odot$ and the mass accretion rate in Eddington unit, $\dot{M} = \dot{m}_b \dot{M}_\mathrm{Edd}$, $\dot{M}_\mathrm{Edd} = L_\mathrm{Edd}/{0.1 c^2} = 1.39\times 10^{18} m_b$~g s$^{-1}$.
Inserting $x_c = 2\times10^3$ and $m_b = 4.9\times 10^8$ to Equation~(\ref{equ:xc}), we obtain $\dot{m}_b = 0.12$, requiring a significantly higher accretion rate compared to that of the typical LLAGNs ($\dot{m}_b=10^{-2}$--$10^{-3}$).
Recently, the equation of the transition radius is modified by the disk evaporation model taking account the effects of viscosity and magnetic field in accretion flow \citep{Qiao:2013}.
In this case,  $x_c = 2\times10^3$ is easily acceptable for $\dot{m}_b \sim 10^{-3}$, which is an expected accretion rate for NGC~4261.  

\begin{figure}
\epsscale{0.8}
\plotone{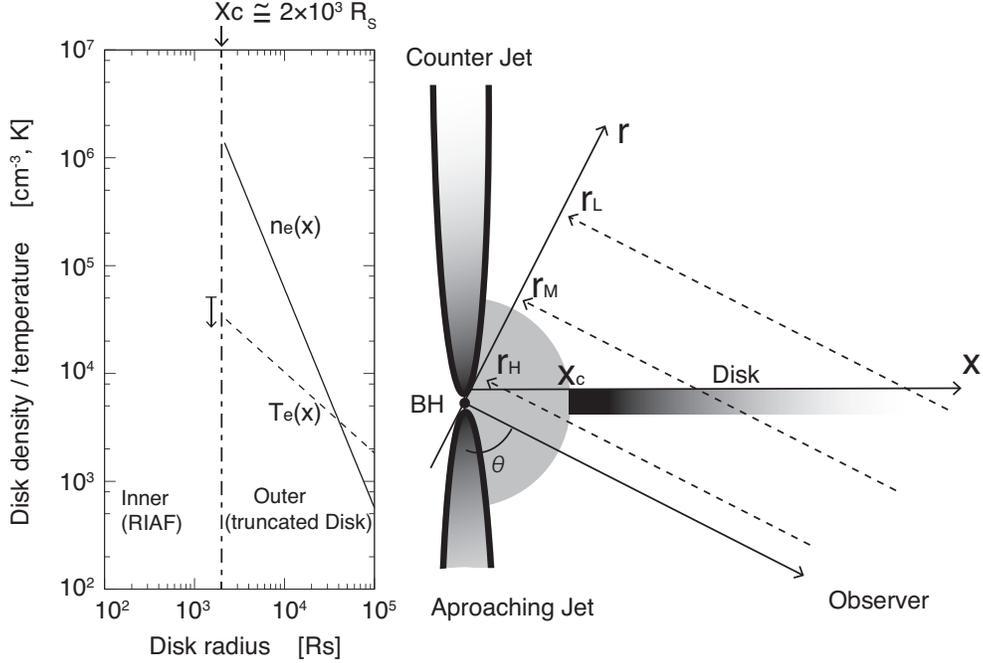}
\caption{
(Left) Density ($n_\mathrm{e}$) and temperature ($T_\mathrm{e}$) profiles of an outer thin accretion disk, assuming $T_\mathrm{e}\propto r^{-0.75}$ and $T_\mathrm{e}=10^4$~K at the area covering the 5~GHz core in the case of $h=0$ and $H_1 =0.01$~pc.  
An arrow denotes the upper limit of $n_\mathrm{e}$ at 22~GHz from $\tau_\mathrm{disk} < 1$, on the basis of the $T_\mathrm{e}$ profile as the outer thin disk, because no deviation from the SSA core shift model was found at 22~GHz (Section~\ref{sec:SSAinjet}).           
(Right) Schematic illustration of our model that explains the observed core shifts in the jets at both sides simultaneously (Section~\ref{sec:diskstructure}).   
The parameters $r_\mathrm{H}$, $r_\mathrm{M}$ and $r_\mathrm{L}$ denotes the observational core position at the high frequency ($\geq$ 22 GHz), the medium frequency (5--15~GHz) and the low frequency ($\leq$ 2.3 GHz) range, respectively.  
At $r_\mathrm{H}$ and $r_\mathrm{L}$, the counter jet is not absorbed sufficiently because of obscuration only by optically thin RIAF at the inner region, and because of an insufficient electron density and/or too low temperature to be ionized at the outer radii.  
On the other hand, at $r_\mathrm{M}$ the counter jet is strongly absorbed by the outer thin disk. 
$x_c$ represent a transition radius.    
}
\label{fig:disk}
\end{figure}

Consequently, this combined model of RIAF and truncated disk is a possible solution to explain the core shift of \objectname{NGC~4261} at both side simultaneously.
Figure~\ref{fig:disk} show this schematic illustration.

\subsubsection{The structure of absorber and the location of X-ray emission}
It is necessary to explain an observed large column density of $N_\mathrm{H} = 6.6 \times 10^{22}\ \mathrm{cm^{-2}}$, where $N_\mathrm{H}$ represents the column density, toward the nucleus \citep[e.g.,][]{Worrall:2010} with an X-ray luminosity of $\sim9\times 10^{40}\ \mathrm{ergs\ s^{-1}}$; the origin of the power-law X-ray emission is controversially associated either with the accretion flow or the base of the jet \citep[e.g.,][]{Zezas:2005,Gliozzi:2003}.

First, we consider the case that the X-ray absorber would be the outer thin disk.  
Substituting $x= 2\times10^3$ in Equation~(\ref{equ:N}), the emission is passing through the region of a density of $n_\mathrm{e}= 2\times10^6$ cm$^{-3}$.  
In case of $L=0.01$ pc, we obtain $N_\mathrm{H} = 6\times10^{22}$ cm$^{-2}$ indicating to be consistent with the X-ray observation.  
However, the geometrically thin disk can obscure only the counter jet in this case.  It is inappropriate to make only the counter jet take responsibility for the X-ray emitting source.

Second, we consider the case that the X-ray absorber would be the inner accretion flow.  We adopt a simple ADAF model \citep{Mahadevan:1997} with $n_\mathrm{e}\propto x^{-1.5}$ as a free-fall density profile and $T_\mathrm{e}\propto x^{-1}$ at $x\ga100$.  The observed X-ray luminosity, suggesting an accretion rate of $\dot{m}_b\sim0.001$, roughly provides an electron temperature and restricts the amount of accreting material.  In this condition, a resultant FFA opacity at $x= 1.5\times10^3$ at 22~GHz is significantly less than unity (cf.~Appendix), which is consistent with our VLBI result.  However, total electron column density toward the nucleus (down to 3~Rs) is at most $N_\mathrm{H}\sim10^{19}\ \mathrm{cm^{-2}}$ through the accretion flow.  Thus, only the accretion flow cannot be responsible for the large column density observed at X-ray.  
Consequently, neither the outer thin disk or the inner RIAF seems to be a sufficient X-ray attenuator; an external absorber, such as a dusty torus, along our line of sight is required to explain X-ray observations.

\section{Conclusion}
We measured the origin of jets at the radio galaxy \objectname{NGC~4261} as the limit of the core position at infinite frequency using a core shift profile. 
The origin is $82\pm16$ $\mu$as (projected) distant from the 43 GHz core position, corresponding to  $310\pm60~R_\mathrm{s}$ (deprojected).  
We also measured the core shift at the counter jet side and confirmed that the core positions at both sides asymptotically converge to the same limit. 
Therefore, we conclude that the jet origin estimated by core shift measurements spatially coincides with the central engine.  

The spectral index maps reveal the possible presence of FFA, $\alpha > 2.5$, in at least a limited area in the counter jet side.    
The core shift profile of the counter jet is not solely attributable to synchrotron self-absorption and requires an additional absorbing component such as a geometrically thin disk.  
Moreover, the core shift and FFA opacity profiles from our result suggest a two-component accretion such as a combined model of RIAF and truncated thin disk that is previously proposed for accretion flows of LLAGNs to explain observed SEDs.  
The density and temperature profiles for the outer thin disk as free--free absorber are estimated to be $T_\mathrm{e}= 1\times10^7 x^{-3/4}$~K and $n_\mathrm{e}= 8\times10^{12} x^{-2}$~cm$^{-3}$, respectively, in the case of $H_1 = 0.01$~pc.  
Neither the outer thin disk or the inner RIAF seems to be a sufficient X-ray attenuator; an external absorber, such as a dusty torus, along our line of sight is required to explain an observed high column density toward the nucleus.

\acknowledgments
The Very Long Baseline Array is operated by the National Radio Astronomy Observatory, a facility of the National Science Foundation, operated under cooperative agreement by Associated Universities, Inc.
This work was partially supported by a Grant-in-Aid for Scientific Research (B; 24340042, AD) from the Japan Society for the Promotion Science (JSPS) and the Center for the Promotion of Integrated Sciences (CPIS) of Sokendai.

%% The displaymath environment will produce the same sort of equation as
%% the equation environment, except that the equation will not be numbered
%% by LaTeX.


%% Included in this acknowledgments section are examples of the
%% AASTeX hypertext markup commands. Use \url without the optional [HREF]
%% argument when you want to print the url directly in the text. Otherwise,
%% use either \url or \anchor, with the HREF as the first argument and the
%% text to be printed in the second.

\clearpage

%% This example uses \plotone to include an EPS file scaled to
%% 80% of its natural size with \epsscale. Its caption
%% has been written to indicate that additional figure parts will be
%% available in the electronic journal.
\clearpage

%\begin{figure}
%\epsscale{0.4}
%\plotone{fig1.eps}
%\caption{
%Contour Image of NGC~4261 at 5 GHz and fitted multi-circular components by {\tt modelfit} in the Difmap. 
%The jet contribution to the image is defined as the difference between this position and that determined from a circular component fit by {\tt JMFIT}
%}
%\label{fig:5G}
%\end{figure}

\clearpage

\begin{figure}
\epsscale{.60}
\plotone{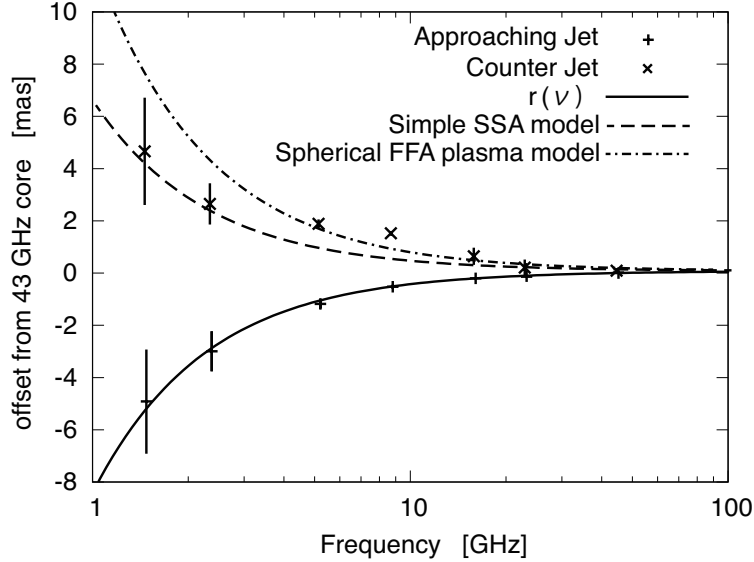}
\caption{
Core shift profile in both side of jet in NGC~4261.
Bars and crosses denote the core positions of the approaching and counter jets, respectively. 
Solid line is a function fitted by the least-square method for the approaching side ($r(\nu)$ in Figure~\ref{fig:cp}). 
At the counter jet side, dashed and dot--dashed lines are predicted core shifts from the SSA model and the spherical FFA plasma model, respectively, assuming symmetry for the two-sided system.   
The parameters $k$ and $b$ equivalent to those of $r(\nu)$ are used for the counter side.  
Given a viewing angle, $\Omega$ at the counter jet side is predicted in both models from the results of the approaching jet side.
}
\label{fig:cs}
\end{figure}

%\begin{figure}
%\epsscale{0.35}
%\plotone{fig6.eps}
%\caption{
%Spectral index maps between adjacent observation frequencies 1~GHz and 2~GHz (top) and 5~GHz and 8~GHz (bottom) synthesized after restoring the images with respect to the lower-frequency beam. 
%In the map of the lower-frequency pair, the region covering  $\pm$ 10 mas shows an inverted spectral index, indicating that absorption occurs over this region. 
%The spectral index $\alpha$ is maximized at 3.5. 
%In the map of the higher-frequency pair, the region of $\alpha >$ 2.5 is located at the east side of the jet (at 0.4 mas), and the spectral index maximum is also positioned on the counter jet.
%}
%\label{fig:spix}
%\end{figure}

\clearpage

\clearpage

%\begin{deluxetable}{ccccccc}
%\tabletypesize{\scriptsize}
%\rotate
%\tablecaption{core positions from 43~GHz core}
%\tablewidth{40pt}
%\tablehead{
%\colhead{} & \colhead{$\Omega$} & \colhead{$\Delta \Omega$}& \colhead{$k$} & \colhead{$\Delta k$} &\colhead{$c$} & \colhead{$\Delta c$}   
%}
%\startdata
%$F_1(\nu)$ & -8.42 & $\pm$ 0.86 & -1.22 & $\pm$ 0.06 & 0.082 & $\pm$ 0.016 \\
%$F_2(\nu)$ &  6.86 & $\pm$ 0.26 & -1.15   & $\pm$ 0.06  & 0.081 & $\pm$ 0.017  
%\enddata
%% Text for table notes should follow after the \enddata but before
%% the \end{deluxetable}. Make sure there is at least one \tablenotemark
%% in the table for each \tablenotetext.
%\tablecomments{Table \ref{tbl-1} is published in its entirety in the 
%electronic edition of the {\it Astrophysical Journal}.  A portion is 
%shown here for guidance regarding its form and content.}
%\tablenotetext{a}{Sample footnote for table~\ref{tbl-1} that was generated
%with the deluxetable environment}
%\tablenotetext{b}{Another sample footnote for table~\ref{tbl-1}}
%\end{deluxetable}
\begin{table}
\begin{center}
\caption{Error budget ($\mu\mathrm{as}$)}
\label{tab:eb}
\begin{tabular}{lccccccc}
\tableline \tableline
Frequency [GHz]                     &1.4      & 2.3   & 5.0  & 8.4    & 15   & 22    & 43   \\
\tableline
Phase-referencing observation &           &         &        &          &           &         &          \\
Beam width/ signal to noise & 52     & 24    &10    & 6       &  4     & 3      & 2       \\
Ionosphere                              &2006 & 779  & 161 & 57    & 17    & 8      & 2       \\
Troposphere                           &13      & 13    & 13   &13     & 13    & 13    &13      \\
Earth orientation                    & 5       &  5      & 5     & 5      & 5       & 5      & 5       \\
Antenna position                   & 2        & 2      & 2     & 2      & 2       & 2      & 2       \\
Apriori source coordinates  & 1        & 1      & 1     & 1      & 1       & 1      & 1       \\
\tableline
Core identification analysis &           &         &        &          &           &         &          \\
Approaching Jet & 349   & 60    & 17   &40     & 3   & 5      & 3        \\
Counter Jet         & 343   & 26    & 5     &50     & 326  & 312 & 19     \\
\tableline 
Total error (RSS)                       &           &         &        &          &           &         &          \\
Approaching Jet & 2055 &792  & 170 & 77   & 30     & 23    &19     \\
Counter Jet         .& 2054 &790  & 169 & 82   & 328  & 313    &25    \\
\tableline
\end{tabular}
\tablecomments{According to the method of \citet{Hada:2011}, error estimation was made for our astrometric measurements in NGC~4261.}
\end{center}
\end{table}

\begin{table}
\begin{center}
\caption{Image performance}
\label{tab:ip}
\begin{tabular}{cccccc}
\tableline \tableline
$\nu$ & $I_\mathrm{peak}$ &$\sigma$            &$\theta_\mathrm{maj}$&$\theta_\mathrm{min}$ & $\theta_\mathrm{p.a.}$\\
(GHz) &  (mJy/beam)             & (mJy/beam)      &  (mas)                            & (mas) & (deg)\\       
(1)       &          (2)                     &               (3)        &         (4)                           & (5)     &(6)\\       
\tableline
1.4      & ~47.4                         & 0.50                   & 8.90                               & 4.45   & 5.5\\
2.3      &~70.1                          & 0.69                   & 5.41                               & 2.60   & 1.1 \\
5.0      &117.1                          & 0.40                   & 2.47                               & 1.19   & 0.06\\ 
8.4      &101.6                          & 0.60                   & 1.51                               & 0.69   & 0.95\\
15       &113.4                          & 0.71                   & 0.85                               & 0.37   & -5.0\\
22       &123.2                          & 1.27                   & 0.55                               & 0.27   & -3.1\\
43       &~72.9                          & 0.94                   & 0.30                               & 0.16   & -2.8 \\
\tableline
\end{tabular}
\tablecomments{Columns are as follows: (1) frequency; (2) peak intensity; (3) Image r.m.s.~noise; (4) major axis size of the synthesized beam; (5) minor axis size of the synthesized beam; (6) position angle of the synthesized beam}
\end{center}
\end{table}

\begin{table}
\begin{center}
\caption{Core positions}
\label{tab:clean}
\begin{tabular}{ccc}
\tableline \tableline
$\nu$ & $r_\mathrm{AJ} $      &$r_\mathrm{CJ}$ \\
(GHz) & (mas)                               & (mas)                         \\ 
(1)       &          (2)                           &     (3)                          \\       
\tableline
1.4      & $-4.92  \pm 2.06$          & $4.74 \pm 2.05$      \\
2.3      & $-2.98  \pm 0.79$          & $2.73 \pm 0.79$      \\
5.0      & $-1.18  \pm 0.17$          & $1.97 \pm 0.17$      \\ 
8.4      & $-0.52  \pm 0.08$           & $1.60 \pm 0.08$     \\
15       & $-0.21  \pm 0.03$          & $0.73 \pm 0.33$     \\
22       & $-0.12  \pm 0.02$           & $0.29 \pm 0.31$      \\
43       &~~~~$0  \pm 0.02$          & $0.17 \pm 0.03$      \\
\tableline
\end{tabular}
\tablecomments{Columns are as follows: (1) frequency; (2) core position in approaching jet with respect to the core at 43 GHz; (3) core position in counter jet with respect to the core at 43 GHz in approaching jet.}
\end{center}
\end{table}

\begin{table}
\begin{center}
\caption{Core shift fitting parameters}
\label{tab:cs}
\begin{tabular}{ccccc}
\tableline \tableline
           &$\Omega$                        &$k$                          & $b$                            &$\chi^2$/n.d.f\\
  (1)    & (2)                                     & (3)                          & (4)                               &  (5 ) \\       
\tableline
$r(\nu)$ & $-8.42 \pm 0.86$ & $1.22 \pm 0.06$ & $0.082 \pm 0.016$ & 0.66/4 \\
\tableline
\end{tabular}
\tablecomments{Columns are as follows: (1) fitting function (Equation~(\ref{equ:cs})); (2)--(4) parameters of core shift fitting ; (5) reduced chi-square. }
\end{center}
\end{table}

\begin{table}
\begin{center}
\caption{Parameters for comparisons}
\label{tab:core43}
\begin{tabular}{cccccccc}
\tableline \tableline
 Object name     &$\Delta r_{43}$         &$M_\mathrm{BH}$ &$\beta$                    & $\theta$                                   & $D$            & $L^{\dagger}_m$     & $L_m$\\
                             &[$R_\mathrm{s}$]     & [10$^8 M_{\odot}$]&                                 &                                                   &                                               & [10$^{40}$ ergs$^{-1}$] &  [10$^{40}$ erg s$^{-1}$]             \\
 (1)                       & (2)                              & (3)                             & (4)                           & (5)                                             & (6)                                         & (7)                               & (8) \\
\tableline
NGC~4261         & $310 \pm 60$         & 4.9$^{a}$                 &0.46$^{d}$             &63$^\circ$$^{d}$                     & 1.1                                         & 3.9$^{h}$ 			        & 2.5\\                              
M~87                    & $23 \pm 6$             & 60$^{b}$                  &0.38--0.62$^{e}$  &15$^\circ$--25$^\circ$$^{g}$& 1.4--1.7                                 & 5.2$^{i}$  			        & 1.4\\ 
BL Lac                  & $10^4-10^6$         &1.7$^{c}$                  &$\geqq0.99^{f}$    &6$^\circ$--10$^\circ$$^{f}$  & 5.6--9.1        &3.6$\times10^{3}$$^{j}$   & 1.0\\                    
\tableline
\end{tabular}
\tablecomments{Columns are as follows: (1) object name; (2) distance of core at 43~GHz from putative central engine; (3) black hole mass; (4) jet velocity in the unit of speed of light; (5) estimated viewing angle; (6) Doppler factor, $D \equiv [\gamma (1-\beta \cos \theta)]^{-1}$; (7) observed luminosity at millimeter wave range; (8) estimated intrinsic luminosity}
\tablenotetext{a}{\citealt{Ferrarese:1996}}
\tablenotetext{b}{\citealt{Gebhardt:2009}}
\tablenotetext{c}{\citealt{Woo:2002}}
\tablenotetext{d}{\citealt{Piner:2001}}
\tablenotetext{e}{calculating from $\beta_\mathrm{app}=0.25$--0.40 \citep{Ly:2007}}
\tablenotetext{f}{\citealt{Jorstad:2005}}
\tablenotetext{g}{\citealt{Acciari:2009}}
\tablenotetext{h}{\citealt{Geldzahler:1981}}
\tablenotetext{i}{\citealt{Davies:2009}}
\tablenotetext{j}{\citealt{Agudo:2010}}
\end{center}
\end{table}

\clearpage

\appendix
%\section{Spherical FFA plasma model}
%This appendix details the methods used to estimate the FFA opacity profile and FFA core shifts. 
%FFA is an alternative core-shifting process to SSA. 
%The spectral index maps at low frequencies, especially, the 1--2~GHz adjacent pair, indicate that the FFA region of $\alpha > 2.5$ extends across not only the counter jet and the core but also the approaching jet. 
%This result implicates FFA as a potential cause of the core shift in the approaching jet. 

\section*{Appendix A \\ Analysis of FFA opacity profile}\label{sec:analysisFFA}
In this section, we estimate the distribution of FFA opacity along the jet axis (from the East to the West). 
We determined an intensity profile along the jet by integrating the flux density of clean components in defined regions, at intervals of 0.25 mas and 1 mas inside and outside of 2 mas from the putative central engine, respectively. 
From the integrated flux densities, we then performed spectral fitting analysis in each area. 
The fitting function for spectral shape $S_\nu$ is described as
\begin{equation}
S_\nu = S_0 \nu ^{\alpha_0} \exp(-\tau_f \nu^{-2.1}), \label{eq:S_nu}
\end{equation}
where 
\begin{equation}
\tau_f \backsimeq 0.46  \left(\frac{n_\mathrm{e}}{1 \mathrm{cm^{-3}}}\right)^2\left(\frac{L}{1 \mathrm{pc}}\right)\left(\frac{T_\mathrm{e}}{1 \mathrm{K}}\right)^{-1.5}.
\label{eq:tau_f}
\end{equation}   

\begin{figure}
\epsscale{0.6}
\plotone{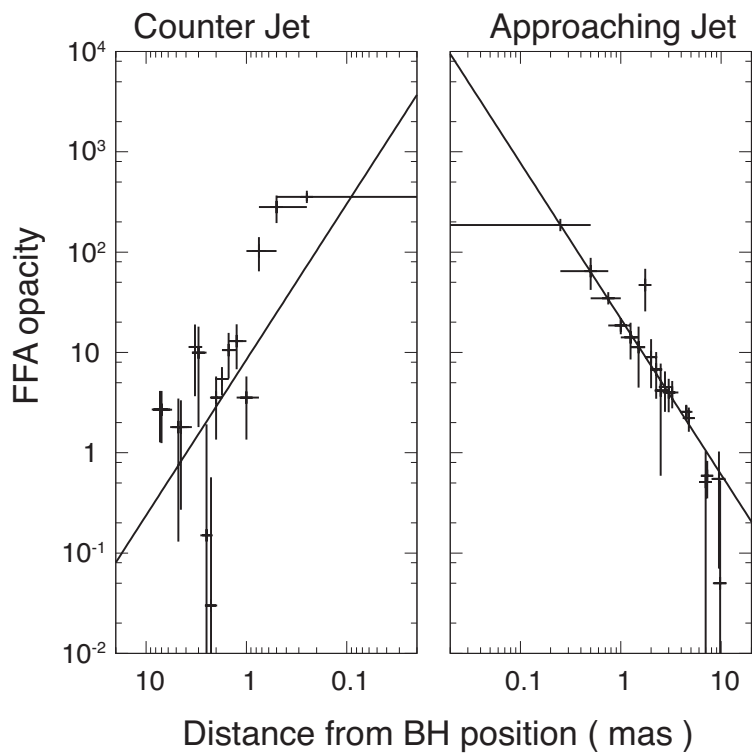}
\caption{
FFA opacity profiles at counter jet (left) and approaching jet (right) sides, assuming that spectral profiles along jets are caused by an FFA process.    
Crosses denote the opacity in each area, and the solid lines denote the least-square fits to the function $\tau_\mathrm{f}=\tau_0 r^{-d}$. 
At counter jet side, we fix $d$ to the same value of the approaching jet side and determined $\tau_0$ by fitting.  
The fitting parameters are given in Table~\ref{tab:tau_ffa}.
}
\label{fig:ffa_tau}
\end{figure}

Here, $S_0$, $\alpha_0$ and $\tau_\mathrm{f}$ are the unabsorbed flux density at 1 GHz, the intrinsic spectral index of synchrotron emission, and FFA opacity, respectively; $n_\mathrm{e}$, $L$ and $T_\mathrm{e}$ are the electron density, path length, and temperature of the plasma, respectively.
Assuming that $\alpha_0$ is fixed at some typical value ($\alpha_0 = -0.7$), $S_0$ and $\tau_\mathrm{f}$ were determined by least-squares fitting. 
Figure~\ref{fig:ffa_tau} shows the determined distribution of FFA opacity $\tau_\mathrm{f}$. 
In addition, we determined the radial profile of FFA opacity at the approaching jet side by fitting to a power-law function; 
\begin{equation}
  \tau_\mathrm{f}  = \tau_0 r ^ {-d}, 
\label{eq:tau}
\end{equation}   
where, $\tau_0$, $d$ are free-fitting parameters. 
The fitted parameters are listed in Table~\ref{tab:tau_ffa}.

\clearpage
\section*{Appendix B \\ Spherical FFA plasma model}\label{Sec:App_ffa}
%\subsection{Core shift due to spherical FFA plasma}\label{Sec:App_ffa}
In this section, we consider an FFA-induced core shift using the radial profile of FFA opacity. 
We assumed that the core position affected by FFA is emitting at the local maximum in the spectral shape. 
The partial derivative of (\ref{eq:S_nu}) with respect to $\nu$ is maximized at $\alpha+2.1\tau =0$. 
Substituting this $\tau$ into (\ref{eq:tau}) and solving the equation for $r$, we obtained the core shift profile at the approaching jet side. 

\begin{figure}
\epsscale{0.6}
\plotone{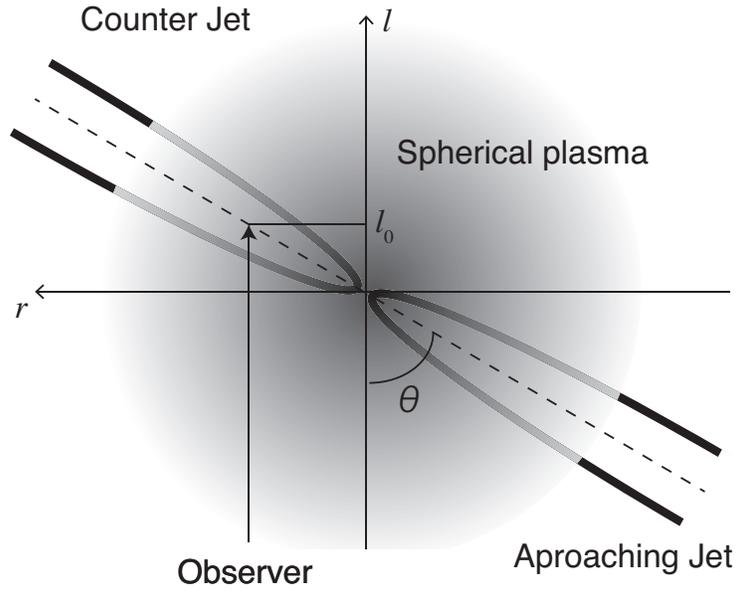}
\caption{
Schematic of spherical FFA plasma model.   
We discuss in the assumptions that radial profiles of electron density and temperature of the plasma are described by power-law functions (Section~\ref{Sec:App_ffa}).  
The counter jet side is more opaque than the approaching jet side at a given radius from the putative black hole for this inclined ($\theta=63\degr$) system of NGC~4261.  
Let the $l$ and $r$ axes be parallel and perpendicular to the line of sight.  
}
\label{fig:ffa_model}
\end{figure}

Assuming the structure of the FFA plasma, the core shift profile at the counter jet side can be derived from the information of the approaching side. 
In this study, we presumed a simple spherical plasma (see Figure~\ref{fig:ffa_model}), which can obscure both the approaching jet and the counter jet. 
If the electron density and temperature of the FFA plasma follow a power-law function, i.e., $n_\mathrm{e}= n_1 x^{-s}$ and $T_\mathrm{e}=T_1 x^{-t}$, the index of the opacity profile, $d$ should be identical in both the approaching and counter sides.
When fitting the opacity profile at counter jet side, we fixed $d$ to the same value as that of the approaching jet and obtained $\tau_0$.
Table~\ref{tab:tau_ffa} provided with this fitting result.
Let us consider the configuration as shown in Figure~\ref{fig:ffa_model}.
Substituting $n_\mathrm{e}$ and $T_\mathrm{e}$ into Equation~(\ref{eq:tau_f}),  we integrate the path length in the plasma along the line of sight from the jet at $l_0= \frac{r}{\tan\theta_0}$.
The radius $x$ from the black hole is described in ($r, l$) coordinate as $x^2 = r^2 + l^2$.
Therefore, the FFA opacity is given by

\begin{equation}
\tau _\mathrm{f}= 0.46 n_1^2 T_1^{3/2} \int_{-\infty}^{l_0} (r^2 + l^2)^{-(s-3t/4)} dl,
\label{eq:int_1}
\end{equation}   

By replacing with $l=\frac{r}{\tan\theta}$, we now convert Equation~(\ref{eq:int_1}) into 
\begin{eqnarray}
\tau _\mathrm{f}&=& 0.46 n_1^2 T_1^{3/2} \int_{0}^{\theta_0} \left\{ r^2 \left(1+ \frac{1}{\tan^{2} \theta}\right) \right\}^{-(s-3t/4)} \frac{r}{\sin^{2} \theta} d\theta, \\
			&=& 0.46 n_1^2 T_1^{3/2} r^{-(2s-3t/2-1)} \int_{0}^{\theta_0} (\sin\theta)^{2s-3t/2-2} d\theta,   \\
			&=& 0.46 n_1^2 T_1^{3/2} r^{-d} \int_{0}^{\theta_0} (\sin\theta)^{d-1} d\theta.  \\
\label{eq:int_2}
\end{eqnarray}

Adopting $d=1.55$ from the analysis of the opacity profile (Table.\ref{tab:tau_ffa}), the result of the integration is obtained in numerical calculations.
An asymmetric profile of the opacity between both of jets is caused by only the different viewing angles, corresponding to the integration ranges $\theta_0$.
Even if $n_1$ and $T_1$ are unknown, we can obtain the opacity profile of the counter jet by using the ratio of the opacity to that of the approaching jet. 

In this way, we can calculate the FFA core shift profile in the counter jet (Figure~\ref{fig:cs}).

\clearpage

\begin{table}
\begin{center}
\caption{Parameters obtained from least-squares fitting of the FFA opacity distribution along R.A.}
\label{tab:tau_ffa}
\begin{tabular}{ccc}
\tableline \tableline
           &$\tau_0$                        &$d$                       \\
(1)      &      (2)                             &(3)                         \\       
\tableline
Approaching Jet & $21.65\pm 0.21$ & 1.55 $\pm$ 0.02    \\
Counter Jet         &  $  8.46\pm 3.56$ &                                   \\
\tableline
\end{tabular}
\tablecomments{Columns are as follows: (1) jet side; (2) FFA opacity at 1 mas ; (3) power-low index for radial dependence.}
\end{center}
\end{table}

%% If you use the table environment, please indicate horizontal rules using
%% Do not put multiple tabular environments within a single table.
%% The optional \label should appear inside the \caption command.

\clearpage

\end{document}